\documentclass[twocolumn,tighten]{aastex63}
\pdfoutput=1
\usepackage{graphicx}
\usepackage{ulem}
\usepackage{amsmath}
\usepackage{wrapfig}
\usepackage{appendix}
\usepackage{enumitem}
\usepackage{multirow}
\usepackage{floatrow}
\usepackage{color}
\usepackage{longtable}
\usepackage{physics}

\newcommand\msun{\, \rm M_\odot}

\newcommand\zsun{\, \rm Z_\odot}

\newcommand\chieff{{\chi_{\rm eff}}}
\newcommand\mbin{{m_{\rm bin}}}
\newcommand\mtrip{{m_{\rm trip}}}
\newcommand\tkl{{T_{\rm KL}}}
\newcommand\jvec{{\mathbf{j}}}
\newcommand\evec{{\mathbf{e}}}
\newcommand\Svec{{\mathbf{S}}}
\newcommand\vesc{{v_{\rm esc}}}
\newcommand\iR{{\mathcal{R}}}

\newcommand\be{\begin{equation}}
\newcommand\ee{\end{equation}}

\def\apjl{ApJL}%
\def\apjs{ApJS}%
\def\aap{A\& A}%
\begin{document}
\shorttitle{Black Hole Mergers in Triples in Clusters}
\shortauthors{Martinez et al.}

\title{Black Hole Mergers from Hierarchical Triples in Dense Star Clusters}

\author[0000-0001-5285-4735]{Miguel A. S. Martinez}
\affil{ Department of Physics \& Astronomy, Northwestern University, Evanston, IL 60208, USA}
\affil{Center for Interdisciplinary Exploration \& Research in Astrophysics (CIERA), Northwestern University, Evanston, IL 60208, USA}
\email{miguelmartinez2025@u.northwestern.edu}

\author[0000-0002-7330-027X]{Giacomo Fragione}
\affil{ Department of Physics \& Astronomy, Northwestern University, Evanston, IL 60208, USA}
\affil{Center for Interdisciplinary Exploration \& Research in Astrophysics (CIERA), Northwestern University, Evanston, IL 60208, USA}

\author[0000-0002-4086-3180]{Kyle Kremer}
\affil{ Department of Physics \& Astronomy, Northwestern University, Evanston, IL 60208, USA}
\affil{Center for Interdisciplinary Exploration \& Research in Astrophysics (CIERA), Northwestern University, Evanston, IL 60208, USA}
\affiliation{TAPIR, California Institute of Technology, Pasadena, CA 91125, USA}
\affil{The Observatories of the Carnegie Institution for Science, Pasadena, CA 91101, USA}

\author[0000-0002-3680-2684]{Sourav Chatterjee}
\affil{Department of Astronomy \& Astrophysics, Tata Institute of Fundamental Research, Homi Bhabha Road, Navy Nagar, Colaba, Mumbai 400005, India}

\author[0000-0003-4175-8881]{Carl L. Rodriguez}
\affil{Astronomy Department, Harvard University, 60 Garden St., Cambridge, MA 02138, USA}

\author[0000-0003-0607-8741]{Johan Samsing}
\affil{Niels Bohr International Academy, The Niels Bohr Institute, Blegdamsvej 17, DK-2100, Copenhagen, Denmark}

\author[0000-0001-9582-881X]{Claire S. Ye}
\affil{ Department of Physics \& Astronomy, Northwestern University, Evanston, IL 60208, USA}
\affil{Center for Interdisciplinary Exploration \& Research in Astrophysics (CIERA), Northwestern University, Evanston, IL 60208, USA}

\author[0000-0002-9660-9085]{Newlin C. Weatherford}
\affil{ Department of Physics \& Astronomy, Northwestern University, Evanston, IL 60208, USA}
\affil{Center for Interdisciplinary Exploration \& Research in Astrophysics (CIERA), Northwestern University, Evanston, IL 60208, USA}

\author[0000-0002-0147-0835]{Michael Zevin}
\affil{ Department of Physics \& Astronomy, Northwestern University, Evanston, IL 60208, USA}
\affil{Center for Interdisciplinary Exploration \& Research in Astrophysics (CIERA), Northwestern University, Evanston, IL 60208, USA}

\author[0000-0002-9802-9279]{Smadar Naoz}
\affil{Department of Physics and Astronomy, University of California, Los Angeles, CA 90095, USA}
\affil{Mani L. Bhaumik Institute for Theoretical Physics, UCLA, Los Angeles, CA 90095, USA}

\author[0000-0002-7132-418X]{Frederic A. Rasio}
\affil{ Department of Physics \& Astronomy, Northwestern University, Evanston, IL 60208, USA}
\affil{Center for Interdisciplinary Exploration \& Research in Astrophysics (CIERA), Northwestern University, Evanston, IL 60208, USA}

\begin{abstract}
Hierarchical triples are expected to be produced by the frequent binary-mediated interactions in the cores of globular clusters. In some of these triples, the tertiary companion can drive the inner binary to merger following large eccentricity oscillations, as a result of the eccentric Kozai-Lidov mechanism. In this paper, we study the dynamics and merger rates of black hole (BH) hierarchical triples, formed via binary--binary encounters in the \texttt{CMC Cluster Catalog}, a suite of cluster simulations with present-day properties representative of the Milky Way's globular clusters. We compare the properties of the mergers from triples to the other merger channels in dense star clusters, and show that triple systems do not produce significant differences in terms of mass and effective spin distribution. However, they represent an important pathway for forming eccentric mergers, which could be detected by LIGO--Virgo/KAGRA (LVK), and future missions such as LISA and DECIGO. We derive a conservative lower limit for the merger rate from this channel of $0.35$  Gpc$^{-3}$yr$^{-1}$ in the local Universe and up to $\sim9\%$ of these events may have a detectable eccentricity at LVK design sensitivity. Additionally, we find that triple systems could play an important role in retaining second-generation BHs, which can later merge again in the core of the host cluster.
\end{abstract}

\section{Introduction}
\label{sect:intro}
Over the past few years, many mergers of binary BHs (BBHs) via gravitational wave (GW) emission have been announced by the LIGO--Virgo collaboration, with an estimated local-universe merger rate of $53.2^{+58.5}_{-28.8}\,\rm{Gpc}^{-3}\rm{yr}^{-1}$ from the O1 and O2 observing runs \citep{abbott+2019a,abbott+2019b}. As the number of publicly announced events increases, most recently with the announcements of the low-mass ratio events GW190412 \citep{LIGO+Virgo2020} and GW190814 \citep{LIGO+Virgo2020b}, it becomes even more crucial to understand the scenarios that lead to the formation of these binaries. There have been many proposed formation channels, both in dense stellar environments and in isolation. These include isolated binary stellar evolution of two massive stars either through common-envelope evolution \citep{dominik+2012,dominik+2013,belczynski+2016,belczynski+2016b,spera2019} or chemically homogeneous evolution of close binaries \citep{demink+mandel2016,mandel+demink2016}, or dynamical assembly in dense stellar environments including young and open star clusters \citep{banerjee2017,banerjee2018a,banerjee2018b,dicarlo+2019,banerjee2020,santoliquido+2020}, globular clusters \citep{portegieszwart+2000,rodriguez+2015,askar+2017,fragione+2018,2018PhRvD..97j3014S, samsdor2018,kremer+2019}, galactic nuclei \citep{oleary+2009,antonini+perets2012,antonini+rasio2016,hamers+2018,hoang+2018,fragrish2019,rasskazov2019,stephan+2019}, and in AGN \citep{stone+2017,bartos+2017,tagawa+2018,mckern2020,tagawa+2019,tagawa+2020}. Another proposed scenario involves primordial BHs merging in the halos of galaxies \citep[e.g.,][]{bird+2016}.

In a stable hierarchical triple, the tidal effect of the tertiary can excite periodic eccentricity oscillations of the inner binary, an effect known as the Kozai-Lidov (KL) mechanism \citep{kozai1962,lidov1962}. A number of studies have expanded on their work to demonstrate the existence of the eccentric Kozai-Lidov (eKL) mechanism and have detailed the potential importance of this phenomenon across many domains of astrophysics \citep[for a review, see][and references therein]{naoz2016}. In particular, binaries excited by the eKL mechanism to high eccentricities can emit GWs more efficiently. As a result, an additional portion of parameter space for mergers is enabled, as wide binaries that would not otherwise merge in isolation are driven to merger by the presence of the tertiary companion. This has previously been shown to enhance the merger rate both in the field and in dynamical environments \citep{kimpson+2016,silsbee+tremaine2017,antonini+2017,rodriguez+antonini2018,grishin+2018,hoang+2018,knight+distefano2019,stephan+2019,liu+lai2019,fragione+bromberg2019,fragk2019,fragione+2019,trani+2019}.

Due to the potential role of triples as GW sources, the possible signatures associated with a triple origin have been under scrutiny. Two of them are particularly relevant. First, eKL-induced mergers may have much higher eccentricities compared to other formation channels \citep{antonini+perets2012,silsbee+tremaine2017,fragrish2019,liu+lai2019,fragk2020}. In an isolated binary, GW emission circularizes the orbit well before the binary reaches the LVK frequency band. However, in a hierarchical triple system, the eKL mechanism can potentially excite the inner binary to arbitrarily high eccentricities such that the inner binary merges before GW emission can circularize the orbit. Second, the spins of the BHs can be used to discriminate between formation channels \citep{liu+lai2017,liu+lai2018,antonini+2018,liu+lai2019,fragk2020}. Specifically, from the inspiral waveform one can extract the effective spin $\chieff$, a weighted measure of the BH spins projected onto the angular momentum vector of the binary orbit. While this quantity is conserved for an isolated binary, it can sweep out a range of values when the spin-orbit misalignment is changing due to De Sitter precession in a triple system. 

Globular clusters (GCs) have been shown to be efficient factories of binary and triple BHs, due to their high central densities. Mass segregation naturally causes the most massive objects to sink into the cluster core, where they assemble hierarchical triple systems through binary-single and binary-binary scatterings \citep{hut+bahcall1983,hut1983,mikkola1983,sigurdsson+phinney1993,rasio+1995,fregeau+2004,ivanova+2008,ivanova2008,antognini+thompson2016,zevin+2019}. While the high density of the cluster core has the effect of rapidly breaking up triples via later dynamical encounters, the formation of BH triples in these environments could still enhance the merger rate of BBHs within GCs \citep{antonini+2016}.

The role of triples in GCs was previously investigated by \citet{antonini+2016}, who showed that triple dynamics enhances the creation of BBH mergers, as well as blue stragglers (through mergers of main-sequence stars). In this study, we will focus solely on the mergers produced by systems composed entirely of BHs. Compared to \citet{antonini+2016}, we use a larger number of GC models, which span a wider range of initial conditions, produced by the \texttt{CMC} code, described in detail by \citet{kremer+2019}. This set of cluster models is of particular interest because it covers the full parameter space of GCs in the Milky Way, so that the triples produced in these simulations are a good representation of the triples dynamically produced in the entire Milky Way GC system \citep{weatherford+2019}. Furthermore, this study also incorporates updated physical prescriptions not present in the study by \citet{antonini+2016}. Specifically, we include updated treatments of post-Newtonian dynamics, which significantly impacts the outcome of both chaotic resonant fewbody encounters \citep{rodriguez+2018,zevin+2019} and single-single encounters \citep{samsing+2019}. As a result, we are able to make direct and self-consistent comparisons between the properties of triple-induced BBH mergers versus other dynamical merger channels. This has implications mainly for the production of mergers with high eccentricities. Additionally, we examine how triples can lead to an increased number of retained second-generation BHs in their host cluster. A larger discussion of the full population of triples can be found in the companion paper by \citet[hereafter Paper I]{PaperI}.

The paper is organized as follows. In \S\ref{sect:background}, we review the secular approximation for hierarchical triples and the relevant modifications that arise from considering relativistic effects and the possible breakdown of the secular approximation. We describe our methods and our sample of triples in \S\ref{sect:methods}, while in \S\ref{sect:results} we present our results. Finally, we summarize our conclusions and discuss the implications of our findings in \S\ref{sect:conc}.

\section{Dynamics of Hierarchical Triples}
\label{sect:background}

We consider a hierarchical triple composed of an inner binary of BHs with masses $m_0$ and $m_1$ and an outer tertiary BH with mass $m_2$. We define $\mbin=m_0+m_1$ and $\mtrip=\mbin+m_2$. We refer to the Keplerian orbital elements $[a,e,i,\omega,\Omega]$ with subscripts ``in'' and ``out'' to refer to the inner and outer binary, respectively, and we define $I$ as their initial mutual inclination. To describe the dynamics of the inner binary, we employ the dimensionless angular momentum vector and eccentricity vector of the inner orbit, $\jvec = \sqrt{1-e_{\rm in}^2}\,\hat{\jvec}$ and $\evec = e_{\rm in} \,\hat{\evec}$, respectively. We also assume that the BHs are born with an initial spin $\Svec_{i} = \chi_{i}(Gm_{i}^2/c)\,\hat{\Svec_{i}}$ where $\chi_i$ is the dimensionless Kerr spin parameter ($i=0,1,2$).

\subsection{Kozai-Lidov Mechanism}
In this study, we consider triple systems that satisfy the stability condition of \citet{mardling+aarseth2001}:
\be
\frac{a_{\rm out}}{a_{\rm in}} > \frac{3.3}{1-e_{\rm out}}\left[ \frac{2}{3} \left( 1+\frac{m_2}{\mbin} \right) \frac{1+e_{\rm out}}{(1-e_{\rm out})^{1/2}} \right]^{2/5} (1- 0.3I/\pi) \,.
\label{eqn:stability}
\ee
If this stability criterion is satisfied, a system is long lived, and thus the Hamiltonian of the system can be described as two separate orbits with a perturbative interaction term \citep{kozai1962,lidov1962}. Assuming a circular outer orbit, one may find the following conserved quantity involving eccentricity, mutual inclination, and argument of pericenter \citep{antognini2015}:
\begin{equation}
C_{\rm KL} = e_{\rm in}^2 \left(1-\frac{5}{2}\sin^2 I \sin^2 \omega_{\rm in} \right)\,.
\label{eqn:cons}
\end{equation}
This conserved quantity has an explicit dependency on the inner eccentricity, inner argument of pericenter, and mutual inclination. As a result, the outer binary induces inclination and eccentricity oscillations as well as pericenter precession. This occurs on a timescale
\be
\tkl\approx \frac{8}{15\pi} \frac{\mbin}{m_2}\frac{P_{\rm out}^2}{P_{\rm in}}(1-e_{\rm out}^2)^{3/2}\,,
\ee
where $P_{\rm in}$ and $P_{\rm out}$ are the orbital periods of the inner and outer orbits, respectively. This process is known as the KL mechanism.\footnote{Note that the double averaged Hamiltonian was presented before Kozai-Lidov, \citep[e.g.,][]{ito+otsuka2019}.} In this level of approximation (the quadrupole approximation) the KL oscillations can only occur within a mutual inclination window of roughly $40^\circ$--$140^\circ$. Allowing for an eccentric outer orbit necessitates an octupole-level approximation. The strength of the octupole-level interaction with respect to the quadrupole interaction can be quantified by 
\be
\epsilon = \frac{m_0-m_1}{m_0+m_1} \frac{a_{\rm in}}{a_{\rm out}} \frac{e_{\rm out}}{1-e_{\rm out}^2} \, ,
\ee
\citep[e.g.,][]{naoz+2013a}.
The inclusion of the octupole terms in the equations of motion allows for much more complex dynamical evolution, such as orbit flips, oscillations from nearly coplanar orbits, and chaotic behavior \citep{naoz+2011,katz+2011,lithwick+naoz2011,naoz+2013a,li+2014}. As this behavior is qualitatively different from the effects presented by Kozai and Lidov, we refer to this as the eKL mechanism. The full octupole equations of motion can be found in \citet{naoz+2013a}.

\subsection{Post-Newtonian Effects}
At 1PN order, relativistic precession will be induced on the inner binary, which happens on a timescale \citep{eggleton+kiseleva2001,naoz+2013,liu+2015,rodriguez+antonini2018}
\be
T_{\rm GR} = (\dv*{\omega_{\rm in}}{t})^{-1} = P_{\rm in} \frac{c^2 a_{\rm in}(1-e_{\rm in}^2)}{G\mbin}\,.
\ee
This precession can quench the eKL oscillations. If $T_{\rm GR} \sim T_{\rm KL}$, then the maximum eccentricity attainable by the inner orbit is limited. If $T_{\rm GR} \ll T_{\rm KL}$, then the KL mechanism is completely suppressed and inner binary eccentricity is unaffected by the tertiary.

At 1.5 PN order, each of the spin vectors $\Svec_0$ and $\Svec_1$ of the inner binary precess due to torques from the inner binary. Thus, it is necessary to include the spin orbit interaction terms \citep{apostolatos+1994}
\be
\dv{\Svec_0}{t} = \frac{2G\mu\nu}{c^2 a_{\rm in} j^3} \left( 1 + \frac{3m_1}{4m_0}\jvec\cross\Svec_0 \right)\,,
\label{eqn:spinorbit}
\ee
where $\mu$ is the reduced mass, $\nu$ is the Keplerian orbital frequency, and $j=|\jvec|$, replacing indices $0\mapsto1$ for the other BH. For a fixed binary, this describes uniform precession around the binary angular momentum vector. In the presence of the KL oscillation, the binary angular momentum will itself precess, thus allowing for a much more interesting behavior. This can be quantified by the evolution of the binary effective spin parameter
\be
\chieff = \frac{m_0\chi_0\cos\theta_0 + m_1\chi_1\cos\theta_1}{\mbin}\,,
\ee
where $\cos\theta_i = \hat{\Svec_i} \cdot \hat{\jvec}$. Though this quantity is conserved for an isolated binary, it evolves over a KL cycle due to the change in the direction of angular momentum. Note that the spin-orbit coupling introduces additional precession on the inner binary, but this depends on the in-plane spin components. Since $\chieff$ only contains information about the spin components perpendicular to the orbital plane, another quantity must be defined with information about the in-plane spin components, known as the effective precession parameter \citep{schmidt+2015}
\be
\chi_{\rm p} = \max \left\{ |\chi_0 \sin\theta_0| , \kappa|\chi_1 \sin\theta_1| \right\}\,,
\ee
where $\kappa = q(4q+3)/(4+3q)$ and $0<q\leq1$ is the mass ratio of the binary. We note that we ignore the backreaction terms, since the binary angular momentum $L\gg S$ during a KL oscillation \citep[e.g.,][]{rodriguez+antonini2018}.

In this study we neglect the 2 PN order effects. Including or discarding these terms in the secular equations of motion do not greatly change the dynamics of the system \citep{naoz+2013}.

Finally, at the 2.5 PN order, the inner binary experiences GW radiation reaction, which is described by the following equations \citep{peters1964}:
\begin{align}
\dv{\evec}{t} \bigg\rvert_{\rm GW} &= -\frac{304}{15} \frac{G^3 m_0 m_1 \mbin}{c^5 a_{\rm in}^4 j^5} \left(1 + \frac{121}{304} e_{\rm in}^2 \right) \evec \label{eqn:peterse}\\
\dv{a_{\rm in}}{t} \bigg\rvert_{\rm GW} &= - \frac{64}{5} \frac{G^3 m_0 m_1 \mbin}{c^5 a_{\rm in}^3 j^7} \left(1 + \frac{73}{24}e_{\rm in}^2 + \frac{37}{96}e_{\rm in}^4 \right) \,. \label{eqn:petersa}
\end{align}
For an isolated binary, these describe a gradual inspiral through the emission of gravitational radiation. The lifetime of such systems until coalescence can be converted to the following integral:
\be
T_{\rm GW} = \frac{12}{19} \frac{c_0^4}{\beta} \int_0^{e_0} \frac{e^{29/19}[1+(121/304)e^2]^{1181/2299}}{(1-e^2)^{3/2}} \dd e\,,
\label{eqn:tpeter}
\ee
where $e_0$ is the initial eccentricity of the binary and the constants in the prefactor are defined as 
\begin{equation*}
c_0 \equiv a_0 e_0^{-12/19} (1-e_0^2) \left[ 1 + \frac{121}{304} e_0^2 \right]^{-870/2299}
\end{equation*}
and 
\begin{equation*}
\beta \equiv \frac{64}{5} \frac{G^3 m_0 m_1 \mbin}{c^5}\,. 
\end{equation*}
For a typical quasi-circular binary composed of two $30\,\msun{}$ BHs to merge within $t_{\rm Hubble}$, the separation must be within  $\sim 0.22\,\mathrm{au}$. However, at extremely high eccentricities, the efficiency of GW radiation is greatly increased, and the merger time greatly reduced. In the case of the example system, for $a_0=0.22\,\mathrm{au}$ and $e_0=0.99$, the time to merger decreases from $t_{\rm Hubble}$ to  $\sim 2.3 \times 10^4\,$yr. For an even more extreme eccentricity of $e_0 = 0.999$, the merger time decreases to just $8\,$yr. While eccentricities this high are usually inaccessible for an isolated compact binary due to circularization during common envelope evolution, the KL mechanism may be able to drive binaries to such eccentricities, depending on the initial conditions.

\section{Methods}
\label{sect:methods}
\subsection{\texttt{CMC}}
We make use of the GC models produced using the \texttt{Cluster Monte Carlo} code. The \texttt{CMC} code is a H\'enon-type Monte Carlo code \citep{henon1971,henon1975} to treat the long term evolution of GCs \citep{joshi+2000,joshi+2001,fregeau+2003,fregeau+rasio2007,chatterjee+2010,umbreit+2012,pattabiraman+2013,morscher+2015,rodriguez+2016,rodriguez+2018,kremer+2019}. This includes detailed treatments of stellar evolution via \texttt{SSE} and \texttt{BSE} \citep{hurley+2000,hurley+2002,chatterjee+2010} with updated prescriptions for compact object formation, two-body relaxation \citep{joshi+2000}, single-single capture \citep{samsing+2019}, three-body binary formation \citep{morscher+2013}, direct stellar collisions \citep{fregeau+rasio2007}, galactic tides \citep{chatterjee+2010,pattabiraman+2013}, and the direct integration of strong 3- and 4-body encounters \citep{fregeau+rasio2007}. Note that in these cluster models, all BHs are assumed to form with no natal spin \citep{fuller+ma2019}. Full details for all the prescriptions used for these \texttt{CMC} models can be found in \citet{kremer+2019}.

In \citet{kremer+2019}, 148 cluster simulations \footnote{The cluster simulations are available for download at \url{https://cmc.ciera.northwestern.edu/home/}} were produced, varying the total number of particles (single stars plus binaries; $N=2\times10^5$, $4\times10^5$, $8\times10^5$, $1.6\times10^6$, and $3.2\times10^6$), initial cluster virial radius ($r_v/\rm{pc}=0.5,\,1,\,2,\,4$), metallicity ($Z/{\rm Z}_\odot=0.01,\,0.1,\,1$), and galactocentric distance ($R_{\rm{gc}}/\rm{kpc}=2,\,8,\,20$). All cluster models initially assume a King potential with King parameter $W_0=5$ \citep{king1962} and a primordial binary fraction of $5\%$. Three of the clusters in the catalog do not form any triples composed entirely of BHs due to collisional runaway before the end of the main sequence lifetime of the most massive stars.

Primordial triples are not included in our models. However, stable hierarchical triples can be formed as the result of strong binary-binary encounters \citep{rasio+1995}. Current limitations of \texttt{CMC} require that triples are broken up into a binary and a single at the end of each integration timestep. Nevertheless, \texttt{CMC} outputs information on the triple, including masses, stellar types, radii, inner and outer semimajor axes and eccentricity, as well as the formation time and properties of the cluster core at formation time. Since we lack information about the mutual orientation of the two orbits, we compensate by creating $10$ different realizations with different mutual orientations \citep{antonini+2016}. In principle, these triple components can be a part of many different triples due to this limitation over the lifetime of the cluster during the simulation.

\subsection{Triples in Clusters}

In our models, we only consider triples produced from binary-binary encounters in the cluster core. We find that the distribution of inner and outer eccentricity for the triples produced by CMC are approximately thermal. Binary-binary encounters can be resonant or non-resonant. In non-resonant encounters, the tighter binary will exchange into the wider binary, with the replaced object kicked out. We find that non-resonant exchange is typical for the BH triples we form in our models. When this happens, according to energy conservation, the newly-formed system will have an outer semimajor axis
\be
a_{\rm out} \simeq a_{\rm wide}\frac{\mbin}{m_{\rm esc}}\,,
\ee
where $a_{\rm wide}$ is the semimajor axis of the wider binary and $m_{\rm esc}$ is the mass of the ejected single \citep{sigurdsson+phinney1993}. For more details, and to see the accuracy of this relation for a wider variety of triple archetypes, see the bottom panel of Fig. 1 and the associated text of Paper I. Since the new semimajor axis depends on this mass ratio, the new orbit can become much wider if the single is, for example, a low-mass star. As a result, even if binaries in the cluster core are very compact, the triple orbits can be very wide. In some cases, the binary-binary interaction can be much more complicated, involving resonant interactions \citep{zevin+2019}. In those cases, it is not straightforward to make such predictions for the endstate of individual systems. For details, see Paper I.

When the single is ejected, the triple will also be imparted a recoil velocity $v_{\rm rec}$. If $v_{\rm rec} > \vesc$, the escape speed from the cluster, the triple will be completely ejected from the cluster. If on the other hand $v_{\rm rec} < \vesc$, the triple's cluster-centric radial orbit will have a new apocenter 
\be
\label{eqn: apo}
r_{\rm apo} \approx r_{\rm c} \sqrt{\frac{v_{\rm esc}^4}{(v_{\rm esc}^2 - v_{\rm rec}^2)^2}-1}\,,
\ee
where $r_{\rm c}$ is the core radius of the cluster. Here, we assume that the encounter occurs at the center of the cluster. This is a reasonable assumption because the potential is approximately constant in the cluster core where these BH triples are formed \citep{sigurdsson+hernquist1993}. While this formula assumes a Plummer potential for the cluster, it is still valid across a wide range of values of $W_0$ for the King models that our clusters assume \citep{antonini+2019}.

In the dense environment of a GC, triples may be perturbed through stellar encounters. To account for this, we calculate for each triple the typical encounter time within the cluster core \citep{ivanova+2008}
\be
\begin{split}
T_{\rm enc} \approx & 8.5 \cross 10^{12} \, {\rm yr} \, P_{\rm out}^{-4/3}M_{\rm tri}^{-2/3}\sigma_{10}^{-1}n_5^{-1} \\
& \times \left[ 1 + 913 \frac{M_{\rm tri} + \langle M \rangle}{2 P_{\rm out}^{2/3}M_{\rm tri}^{1/3}\sigma_{10}^2} \right]\,,
\end{split}
\label{eqn:tenc}
\ee
where $P_{\rm out}$ is the outer binary period in days, $\mtrip$ is the total mass of the triple in $\msun$, $\sigma_{10}$ is the central velocity dispersion of the cluster in units of 10 $\rm{km}$ $\rm{s^{-1}}$, $\langle M \rangle$ is the average mass of an object in the cluster in $\mathrm{M_\odot}$, and $n_5$ is the number density of objects in units of $10^5$ $\rm{pc}^{-3}$. In order to account for the recoil of triples into orbits that may extend outside the core, we can replace the quantities in Eq. \ref{eqn:tenc} describing the cluster with their local versions, which are monotonically declining functions of $r$ \citep{king1962}. Then we can average as follows in order to account for the triple's total orbit:
\be
\langle T_{\rm enc} \rangle = \int_0^{r_{\rm apo}} r^2 T_{\rm enc}(r) \dd r\,.
\ee
In reality, this approach does not take into account further evolution of the orbit due to dynamical friction, though it is sufficient since mergers due to eKL will typically happen before significant evolution can take place. For a triple that is completely ejected from the cluster, we treat the encounter time as infinite.

\subsection{The Triple Sample}

We create $10$ different realizations of the orbital orientations, sampling the mutual inclination from an isotropic distribution, i.e., uniform in $\cos I$ between $0$ and $1$ and the other two Euler angles uniformly between $0$ and $2\pi$. Finally, we sample the recoil velocities of each triple $10$ times. The procedure for this is described in detail in Paper I. Thus, we create for each triple a total of $100$ different realizations.

Of the triples in the cluster catalog, $63,508$ are composed of three BHs. A small number of the triples are only stable at certain eccentricities, so after resampling each triple $100$ times, we once again evaluate their stability using Eq. \ref{eqn:stability}, finding a total sample size of $6,090,030$. Those that remain stable are integrated forward in time numerically using the publicly available secular code \texttt{Kozai}\footnote{We use the C++ version of this code available at https://github.com/carlrodriguez/kozai/tree/master/. We also tested a subpopulation of the sample using the secular code \texttt{OSPE} \citep{naoz+2013a}.} \citep{rodriguez+antonini2018,antonini+2018} until either the integration time reaches $\min (t_{\rm Hubble} - t_{\rm formation},\,1000 \tkl,\,\langle T_{\rm enc} \rangle)$ or the triple reached the LVK frequency band $f_{\rm GW} = 10 \, \mathrm{Hz}$, producing a merger at time $T_{\rm merger}$. We also keep track of the eccentricity of the inner binary when the peak frequency passes through the values $0.01$ Hz, $1$ Hz, and $10$, characteristic frequencies for the LISA, DECIGO, and LVK detectors, respectively. We calculate the frequency as the highest harmonic produced by the inner binary as defined by \citet{wen2003}:
\be
f_{\rm GW} = \frac{\sqrt{G\mbin}}{\pi} \frac{(1 + e_{\rm in})^{1.1954}}{[a_{\rm in} (1-e_{\rm in}^2)]^{1.5}}\,.
\label{eqn: GWfreq}
\ee

\section{Results}
\label{sect:results}

\begin{figure}
\includegraphics[width=0.96\textwidth]{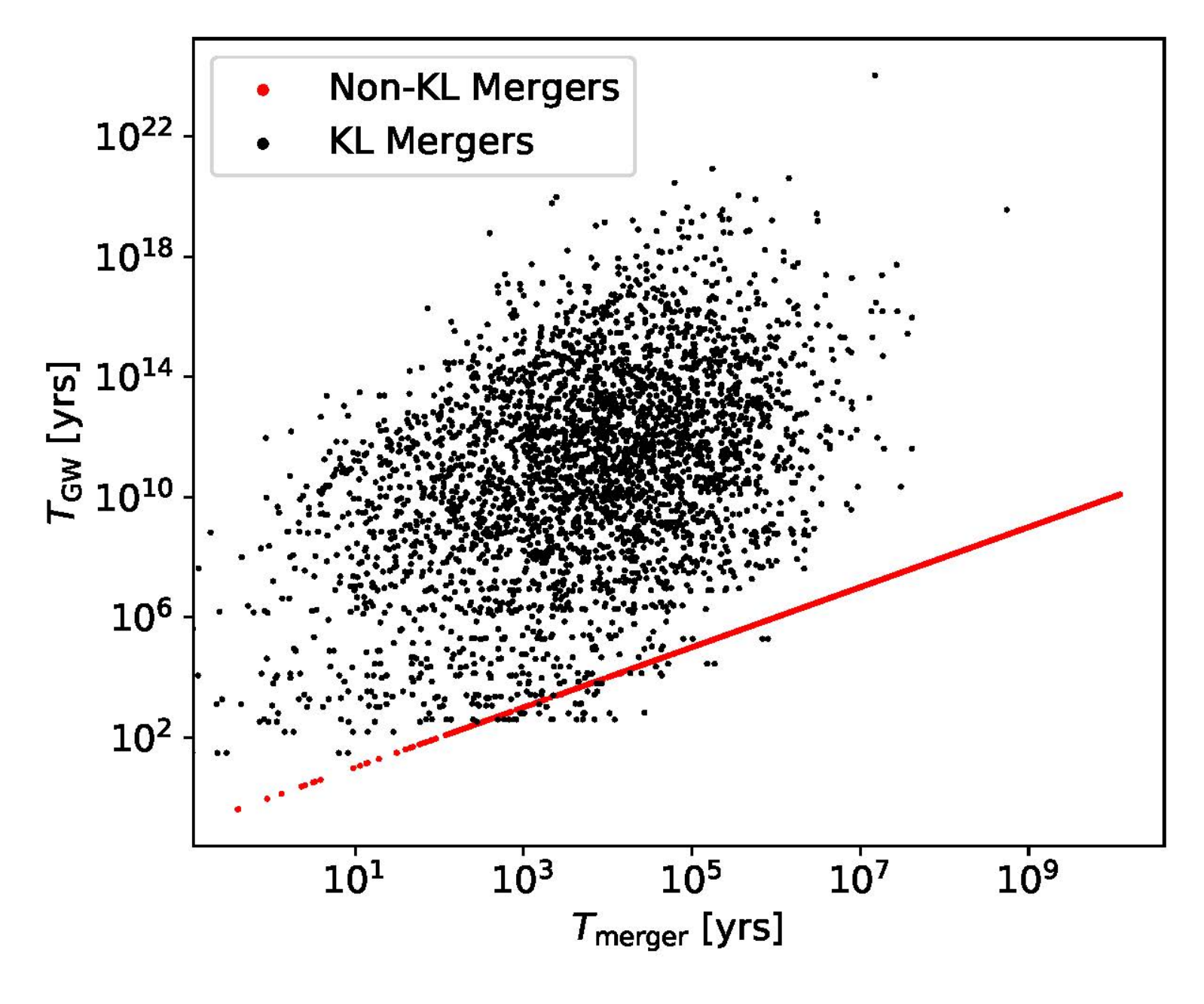}
\caption{Merger time versus GW inspiral time (Eq. \ref{eqn:tpeter}) if the inner binary were isolated for merging triples. Two populations clearly emerge from this comparison. Where $T_{\rm merger} \approx T_{\rm GW}$ ($N=55,246$), the tertiary does not influence the motion of the inner binary. These non-KL mergers are shown in red. KL mergers where the tertiary does impact the merger time such that $T_{\rm merger} \not\approx T_{\rm GW}$ ($N=30,783$) are shown in black.}
\label{fig:timescales}
\end{figure}

\begin{figure}
\includegraphics[width=0.96\textwidth]{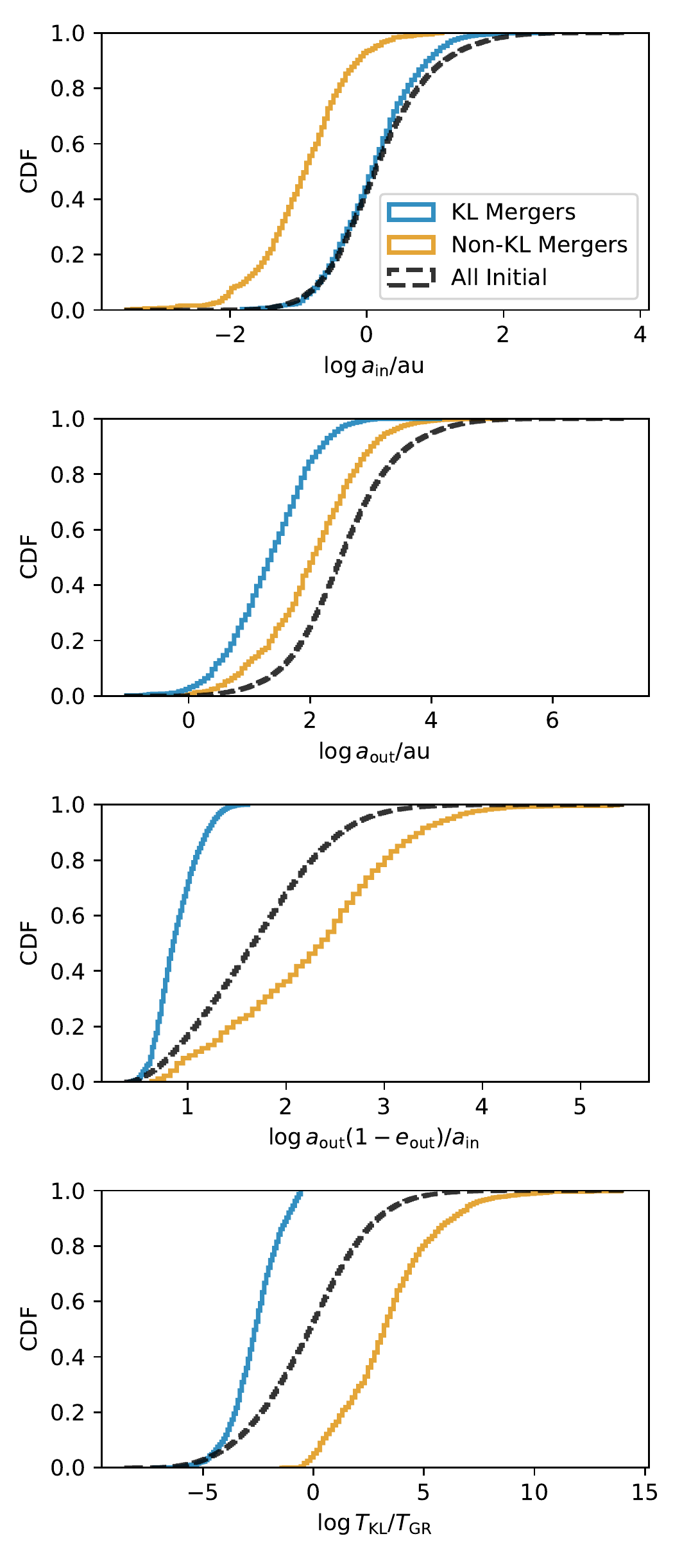}
\caption{Cumulative distribution functions (CDFs) of initial orbital properties of merging triples. Top: inner binary semimajor axis. Center Top: outer binary semimajor axis. Center Bottom: ratio of outer binary pericenter to inner binary semimajor axis. Bottom: ratio of $T_{\rm KL}$ to $T_{\rm GR}$. KL mergers are shown in blue and Non-KL systems are shown in orange. The full initial conditions are shown in black.}
\label{fig:orbits}
\end{figure}

\begin{figure}
\includegraphics[width=0.96\textwidth]{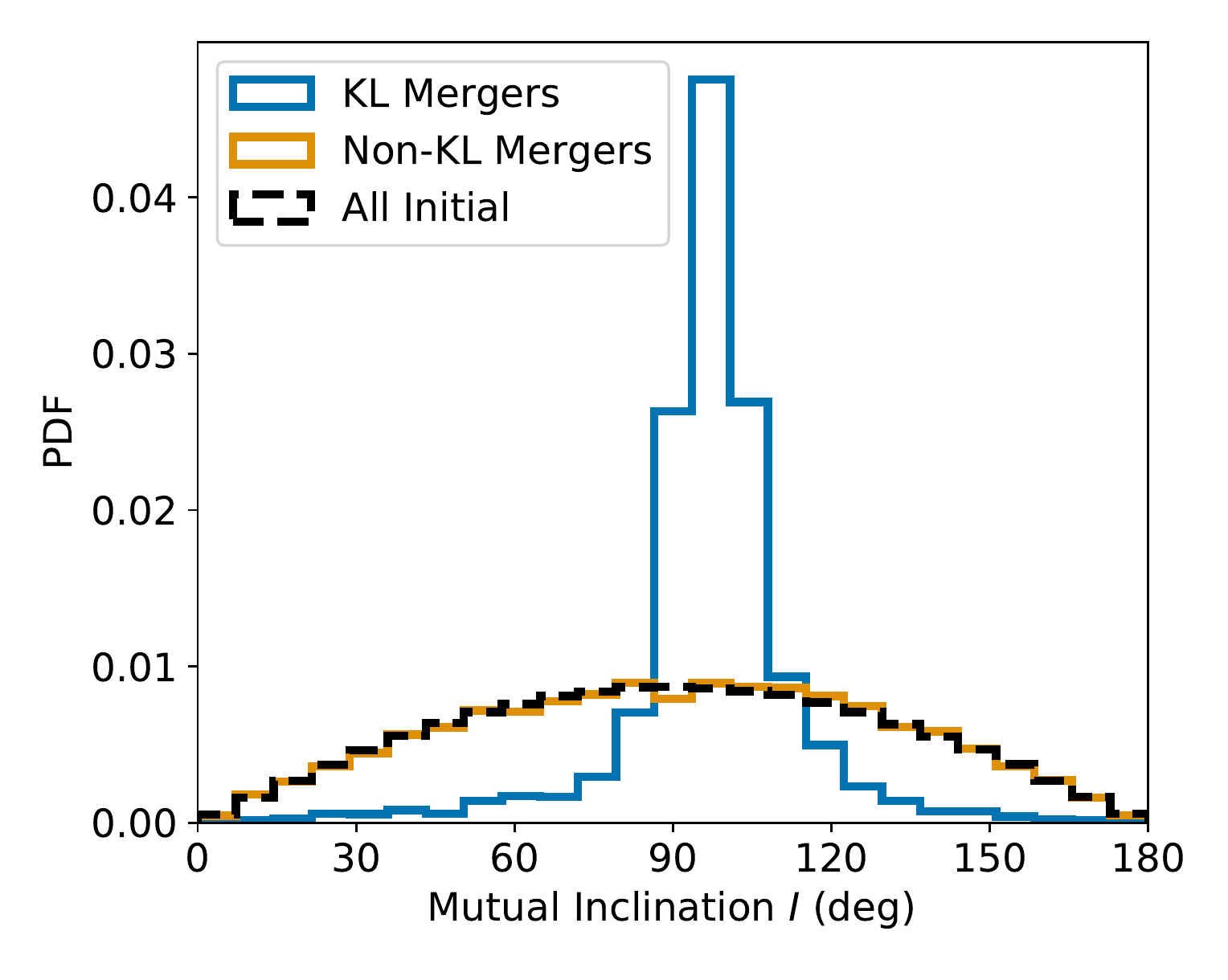}
\caption{Initial inclination of merging triples. KL mergers are shown in blue. Non-KL systems are shown in orange. The full initial conditions are shown in black. While the non-KL systems show no preference to orientation, the KL systems show a clear preference for nearly-coplanar orientations.}
\label{fig:incs}
\end{figure}

\begin{table}
    \centering
    \begin{tabular}{c|c|c|c}
        \hline
        $\mathbf{N}$ ($\mathbf{10^5}$) & $\mathbf{N_{\rm sims}}$ &  $\mathbf{\langle N_{\rm KL}\rangle}$ & $\mathbf{\langle N_{\rm non-KL}\rangle}$\\
        \hline
        $2$  &  $36$  &  $0.50$  &  $0.73$\\
        $4$  &  $36$  &  $1.35$  &  $2.24$\\
        $8$  &  $36$  &  $2.33$  &  $4.63$\\
        $16$  &  $33$  &  $4.14$  &  $7.22$\\
        $32$  &  $4$  &  $5.16$  &  $10.16$\\
        \hline
        \hline
       $\mathbf{r_{\rm v}}$ (\textbf{pc})  & $\mathbf{N_{\rm sims}}$ &  $\mathbf{\langle N_{\rm KL}\rangle}$ & $\mathbf{\langle N_{\rm non-KL}\rangle}$\\
        \hline
        $0.5$  &  $33$  &  $2.16$  &  $5.76$\\
        $1$  &  $38$  &  $2.73$  &  $4.53$\\
        $2$  &  $38$  &  $2.22$  &  $3.63$\\
        $4$  &  $36$  &  $1.35$  &  $1.45$\\
        \hline
        \hline
        $\mathbf{Z}$ ($\mathbf{{\rm Z}_\odot}$) & $\mathbf{N_{\rm sims}}$ &  $\mathbf{\langle N_{\rm KL}\rangle}$ & $\mathbf{\langle N_{\rm non-KL}\rangle}$\\
        \hline
        $0.01$  &  $47$  &  $2.23$  &  $3.44$\\
        $0.1$  &  $50$  &  $2.24$  &  $3.76$\\
        $1$  &  $48$  &  $1.90$  &  $4.22$\\
        \hline
    \end{tabular}
    \caption{Average number of mergers sorted by cluster $Z$, $N$, and $r_{\rm v}$. We divide the number of mergers from clusters with a given property by the number of clusters with the given property $N_{\rm sim}$. Three cluster simulations with $r_{\rm v}=0.5$ pc, $Z=0.01\zsun$, and $N=1.6\cross10^6$ but different $r_{\rm g}$ were halted due to collisional runaway and produced no triples, and thus are not included in the total $N_{\rm sim}$. Note that these mergers are from a resampled population, so to get the true number of mergers from a given cluster, the number of mergers has been divided by $100$.}
    \label{tab:mergers}
\end{table}

From the secular integrations, we find that $86,029$ ($1.4\%$) of the systems merge before $\langle T_{\rm enc} \rangle$.
In Figure \ref{fig:timescales}, we compare $T_{\rm GW}$ to $T_{\rm merger}$ for all the merging systems and find the emergence of two distinct populations. In $55,246$ of the mergers, or $64\%$, $T_{\rm GW} \approx T_{\rm merger}$ and the evolution is completely dominated by GW radiation from the outset, such that the presence of the tertiary has a negligible effect (hereafter non-KL mergers). On the other hand, $30,783$ ($36\%$) of the total mergers are eKL assisted, wherein the GW emission only happens after the eccentricity of the inner binary is excited to very large values due to the presence of the tertiary (hereafter KL mergers). 

We can obtain upper and lower bounds on the number of mergers by considering the most pessimistic and optimistic recoil velocities. By doing so, we find that the number of non-KL mergers can be as low as $26620$ and as high as $205300$. On the other hand, the number of KL mergers remains essentially the same, with lower and upper bounds of $30520$ and $32280$, respectively. The number of KL mergers does not change very much since eKL-induced mergers typically happen on very short timescales $\lesssim10^5$ yr, so that the survival of the triple prior to merger is not very sensitive to changes in the encounter time. On the other hand, the merger time due solely to GW radiation is extremely sensitive to the initial inner semimajor axis, so the number of non-KL mergers is extremely sensitive to different values of $\langle T_{\rm enc} \rangle$.

The differences between these two populations are clear from Figure \ref{fig:orbits}. The non-KL triples in general have much more compact inner orbits and larger outer orbits; the top panel shows that  $\sim 50\%$ of the non-KL systems have $a_{\rm in}\lesssim 0.1 \,\mathrm{au}$, whereas this is true for only $\sim 3\%$ of the KL induced merging systems. On the other hand,  $\sim 50\%$ of the non-KL systems have $a_{\rm out}\gtrsim 100 \,\mathrm{au}$, while this is only true for $\sim 15\%$ of the KL systems. As a result, $\sim 50\%$ of the KL systems have a ratio of less than $\sim 10$ between the outer orbit pericenter and the inner orbit semimajor axis and all of them have a ratio of less than $\sim 30$ between these two distances. On the other hand,  $\sim 40\%$ of the non-KL systems have ratios above $\sim 100$ and  $\sim 20$ have ratios above $\sim 1000$. Since $T_{\rm KL}\propto a_{\rm out}^3/a_{\rm in}^{3/2}$, the non-KL systems have extremely large KL timescales compared to the 1PN precession timescale and so the KL oscillation is suppressed. We can see that these ratio distributions manifest themselves in the final panel comparing $T_{\rm KL}$ and $T_{\rm GR}$. While the initial sample spans values of $T_{\rm KL}/T_{\rm GR}$ from $\sim 10^{-6}$ to $\sim 10^6$, all of the KL merging systems have $T_{\rm KL}/T_{\rm GR}\lesssim 1$ while the opposite is true for all but $\sim 2\%$ of the non-KL merging systems.

Figure \ref{fig:incs} reinforces this interpretation, where we show the initial mutual inclination of the triple systems that lead to a merger in the inner binary. The initial mutual inclination of the non-KL systems are oriented isotropically, while the KL systems have initial mutual inclinations peaked near $100^\circ$. The eKL mechanism causes larger eccentricity excitations in the inner binary for more highly-inclined systems, so naturally the majority of the merging systems will be initially near-perpendicular \citep[e.g.][]{naoz2016}. This peak does not occur at $90^\circ$ because of symmetry breaking in the quadrupole order expansion when relaxing the test-particle approximation \citep[e.g. Eq. 63 of ][]{liu+2015}. On the other hand, since the evolution of the non-KL systems is completely dominated by GW emission, the initial relative inclination of the inner and outer orbit is irrelevant, thus leading to an isotropic distribution.

In Table~\ref{tab:mergers}, we consider the dependence of the merger number on the cluster properties, namely $Z$, initial $r_{\rm v}$, and initial $N$. We divide the total number of mergers that took place in a cluster with a given property in order to compute an average merger number. As Paper I showed, larger initial $N$, smaller initial $r_{\rm v}$, and lower $Z$ promote the efficient creation of triples due to promoting higher stellar densities and increasing the number of encounters. This is reflected in the average merger numbers that we obtain. However, note that if stellar densities become too high, such as in the $r_{\rm v}=0.5$ pc case, these higher stellar densities cause triples to be reprocessed (or even disrupted) by other cluster members before they are able to produce mergers.

\subsection{Eccentricity}

\begin{figure*}
\centering
\includegraphics[width=0.85\textwidth]{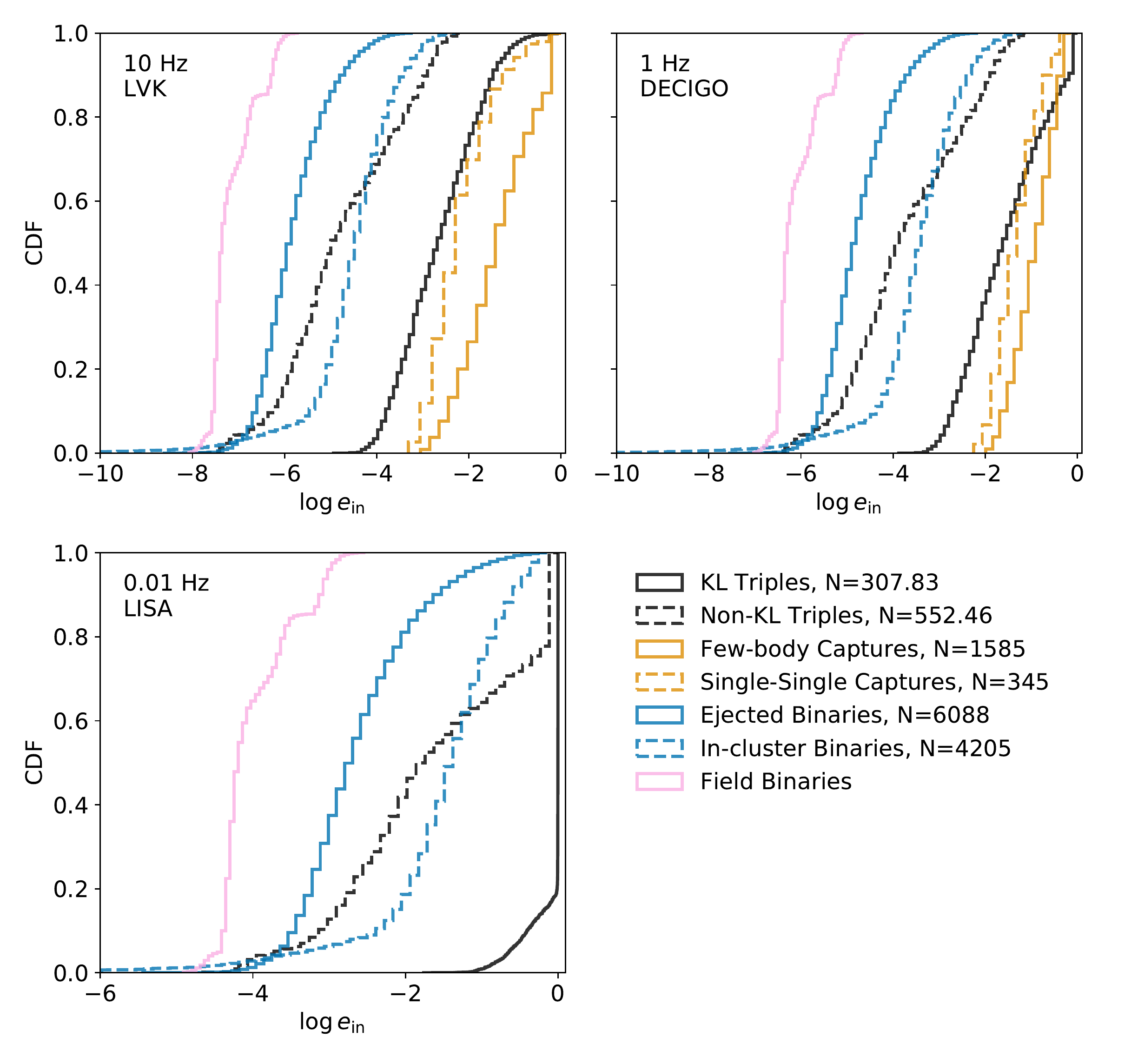}
\caption{Eccentricity of merging binaries by triple and binary evolution at various frequencies. Displayed are the eccentricity at $10$ Hz (top left), $1$ Hz (top right), and $0.01$ Hz (bottom). In the case of $0.01$ Hz, since the eccentricity oscillates due to the KL mechanism, the peak frequency can pass through this value many times over the course of its evolution. When this happens, only the eccentricity at the final passage through this value is shown. In addition to the KL mergers (solid black) and non-KL mergers (dashed black), also included are the few-body capture (solid orange), single-single capture (dashed orange), ejected (solid blue), and in-cluster (dashed blue) binary mergers from \citet{kremer+2019}, along with results from \citet{kremer+2019eccs} which show eccentricities for field binaries (solid pink). As the few-body capture and single-single capture mergers form at frequencies above $0.01$ Hz, they are not included in the bottom panel.}
\label{fig:eccs}
\end{figure*}

In Figure \ref{fig:eccs}, we show the eccentricity spectrum produced by different populations of merging binaries at a range of frequencies representing the peak frequencies of LVK ($10$ Hz), DECIGO ($1$ Hz), and LISA ($0.01$ Hz). In addition to the KL mergers and non-KL mergers from this study, we also show the binary mergers from the same cluster models, subdivided into four different categories \citep[e.g.][]{samsdor2018, zevin+2019, kremer+2019}. The ejected mergers are mergers of binaries ejected from the cluster \citep{portegieszwart+2000}. In-cluster mergers are defined as mergers of binaries formed by dynamical encounters that occur within the cluster \citep{rodriguez+2018}. Few-body captures are the mergers that occur during a resonant 3- or 4-body encounter \citep{samsing+2014, zevin+2019}. Finally, single-single captures occur when two BHs on initially hyperbolic orbits come sufficiently close for GW radiation to create a binary \citep{samsing+2019}. We also include a population of field binaries from \citet{kremer+2019eccs} which in turn were computed using \texttt{COSMIC} \citep{breivik+2019}.

Note that the capture merger channels form binaries with initially high frequency. In order to compute the formation frequency for each binary, we follow the procedure outlined in detail in \citet{zevin+2019}. In short, we use Eq.~\ref{eqn: GWfreq} and
\be
\left\langle \dv{a_{\rm in}}{e_{\rm in}} \right\rangle = \frac{12}{19}\frac{a_{\rm in}}{e_{\rm in}}\frac{[1+(73/24)e_{\rm in}^2+(37/96)e_{\rm in}^4]}{(1-e_{\rm in}^2)[1+(121/304)e_{\rm in}^2]}\,,
\label{eqn:petersea}
\ee
which is found from the coupled Eqs.~\ref{eqn:peterse}--\ref{eqn:petersa} in \citet{peters1964}. These equations are not differentiable at a given frequency, $f_{\rm GW}$, when they form above that frequency, as $e(f_{\rm GW})>1$. Therefore, we compute the pericenter distance $R_p$ at a reference formation eccentricity $1-e=10^{-3}$ and compare this to the semimajor axis $a_{\rm circ}$ of the binary if it were on a circular orbit with frequency $f_{\rm orb} = f_{\rm GW}/2$. If $R_p[f_{\rm GW}]<a_{\rm circ}[f_{\rm orb}]$, then the binary formed above $f_{GW}$. As a result, the LISA panel of this figure does not show the lines representing the latter two of these four categories as these mergers form with initial frequencies above $0.01$ Hz. Similarly, we exclude mergers which form above $1$ Hz in the DECIGO panel. However, in the LVK panel, we assign mergers which form above $10$ Hz $e_{\rm in}=0.999$.

The top left panel shows that the KL mechanism produces much more eccentric mergers in the LVK frequency band compared to in-cluster and ejected mergers, but also shows that the KL mergers are in general less eccentric than the few-body and single-single captures. All of the non-capture binary-mediated mergers have eccentricity less than $\sim10^{-3}$. The non-KL triple merger eccentricity distribution reflects the fact that it is a subset of the standard ejected and in-cluster binary merger channels. On the other hand, $\sim 30\%$ of the KL mergers have eccentricity above $10^{-2}$ and a few percent are in excess of $\sim 0.1$. This is consistent with previous results from \citet{antonini+2016}. However, when compared to the few-body and GW capture scenarios, it is clear that the contribution of eccentric sources by KL mergers will be subdominant. In particular, almost $\sim 20\%$ of the few-body mergers produced during resonant interactions were formed over $\sim 10$ Hz.

When using a secular code to evaluate KL-driven evolution, there are many cases for which the secular equations of motion cannot accurately reproduce the evolution of the system during the peaks of the KL oscillation. In addition to producing more mergers with shorter merger times, direct $N$-body integration produces a much higher peak in the eccentricity spectrum than when using a secular code \citep{antonini+2016,grishin+2018,fragione+2019trip}. This is because the inner binary angular momentum can become arbitrarily small during the peak of the KL oscillation, such that GW emission only begins to dominate the evolution once the inner binary frequency is near or within the LVK frequency band. Note that this would only affect the shape of the solid black curve showing the KL merging systems because the other populations do not experience the high eccentricity oscillations that would allow this to happen. We discuss this further in \S\ref{subsect:n-body}.

From the bottom panel, it is clear that all the KL mergers have high eccentricities at $f_{\rm GW}\sim 0.01$ Hz. More than $\sim 80\%$ of these binaries have eccentricities in excess of $\sim 0.9$. Remarkably, however, KL mergers will potentially enter and leave the LISA sensitivity range many times before merger, whereas all the other merger channels will evolve with monotonically increasing frequency. \citet{randall+xianyu2019}, \citet{hoang+2019}, and \citet{emami+loeb2020} have shown that it is possible for LISA to directly detect the eccentricity oscillations of hierarchical triples, whether the tertiary is a third stellar-mass body or an SMBH.\footnote{\citet{deme+2020} has also investigated this in the context of a stellar-mass BH orbiting an IMBH perturbed by an SMBH.} This is only possible when they are close to the peak of the eccentricity oscillation, when the rapid change in eccentricity results in a rapid change in the emitted peak harmonic frequency. Both works find that the typical timescale for changes in the characteristic strain will happen over the timescale of $\mathcal{O}(10^2)$ days. \citet{emami+loeb2020} showed that these methods can detect BH triples with stellar-mass components as far away as M87. While there could in principle be many Galactic triples undergoing KL oscillation, the population of triples detectable in this way will be dominated by triples in the Galactic field. We note that while the bottom panels show that $\sim 20\%$ of non-KL triples have $e_{\rm in}>0.99$, these sources will not show the same increase and subsequent decrease in eccentricity since the evolution is dominated by GW emission.

In order to have a better estimate for the number of eccentric sources and track their evolution prior to merger, it is necessary to probe the sub-Hz frequencies between the frequency ranges of LVK and LISA with future decihertz range detectors such as DECIGO \citep{kawamura+2011}. We find that  $\sim 20\%$ of the KL-driven mergers will have $e_{\rm in}>0.1$ in the DECIGO band, comparable to the fraction of few-body captures and single-single captures that have $e_{\rm in}>0.1$. Moreover, since some sources will form within the DECIGO frequency range, it will be possible to further distinguish the nature of different sources. However, as the absolute number of few-body capture and single-single capture mergers from a given cluster is higher than the number of KL mergers, we expect that the contribution of the KL mergers to the eccentric merger rate will be subdominant. We once again reiterate that the secular code underestimates the peak of the eccentricity spectrum, and thus we expect that a significant fraction of the KL-driven mergers will have high eccentricity in the DECIGO band.

These results are of particular interest in light of the recent detection of GW190521, which has component masses consistent with a dynamical origin \citep{GW190521a,GW190521b}. \citet{romero-shaw+2020} have argued that the waveform could be consistent with a moderately eccentric binary at $10$ Hz, depending on the priors used. This detection emphasizes the necessity of quantifying the contributions of the various dynamical channels to the population of eccentric merger events.

\subsection{Masses}

\begin{figure*}
\centering
\includegraphics[width=0.96\textwidth]{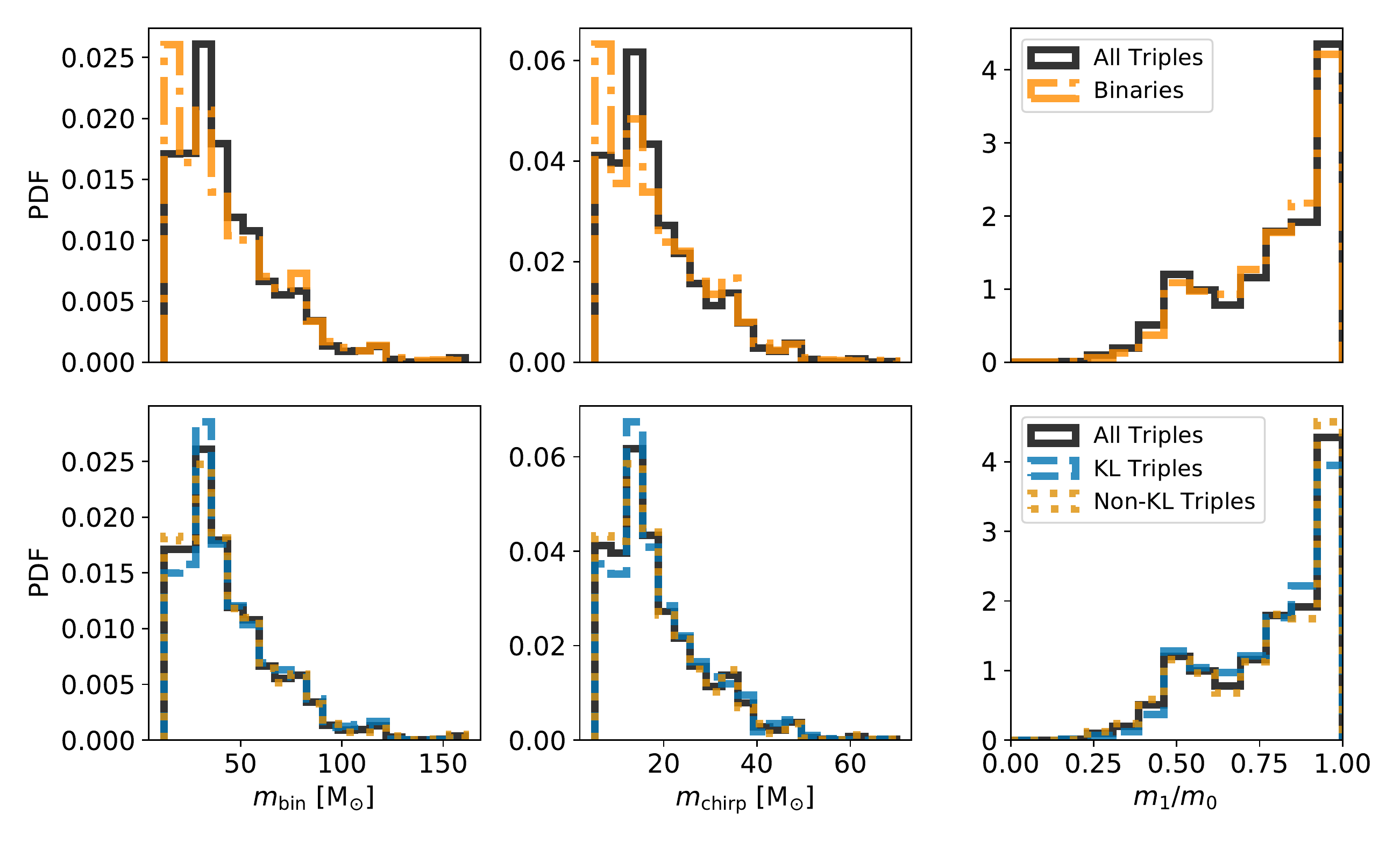}
\caption{Distributions of total binary mass (left), chirp mass (middle), and mass ratios (right) of merging binaries. In the top row, the combined triple population (solid black) and the binary merger population (dash-dotted orange) from the cluster models of \citet{kremer+2019} are compared. In the bottom row, the KL mergers (dashed blue), the non-KL mergers (dotted orange), and the combined triple population (solid black) are compared. The slight differences in the distributions underscore that triple evolution has at most a second-order effect on mass.}
\label{fig:masseshist}
\end{figure*}

\begin{figure*}
\centering
\includegraphics[width=0.96\textwidth]{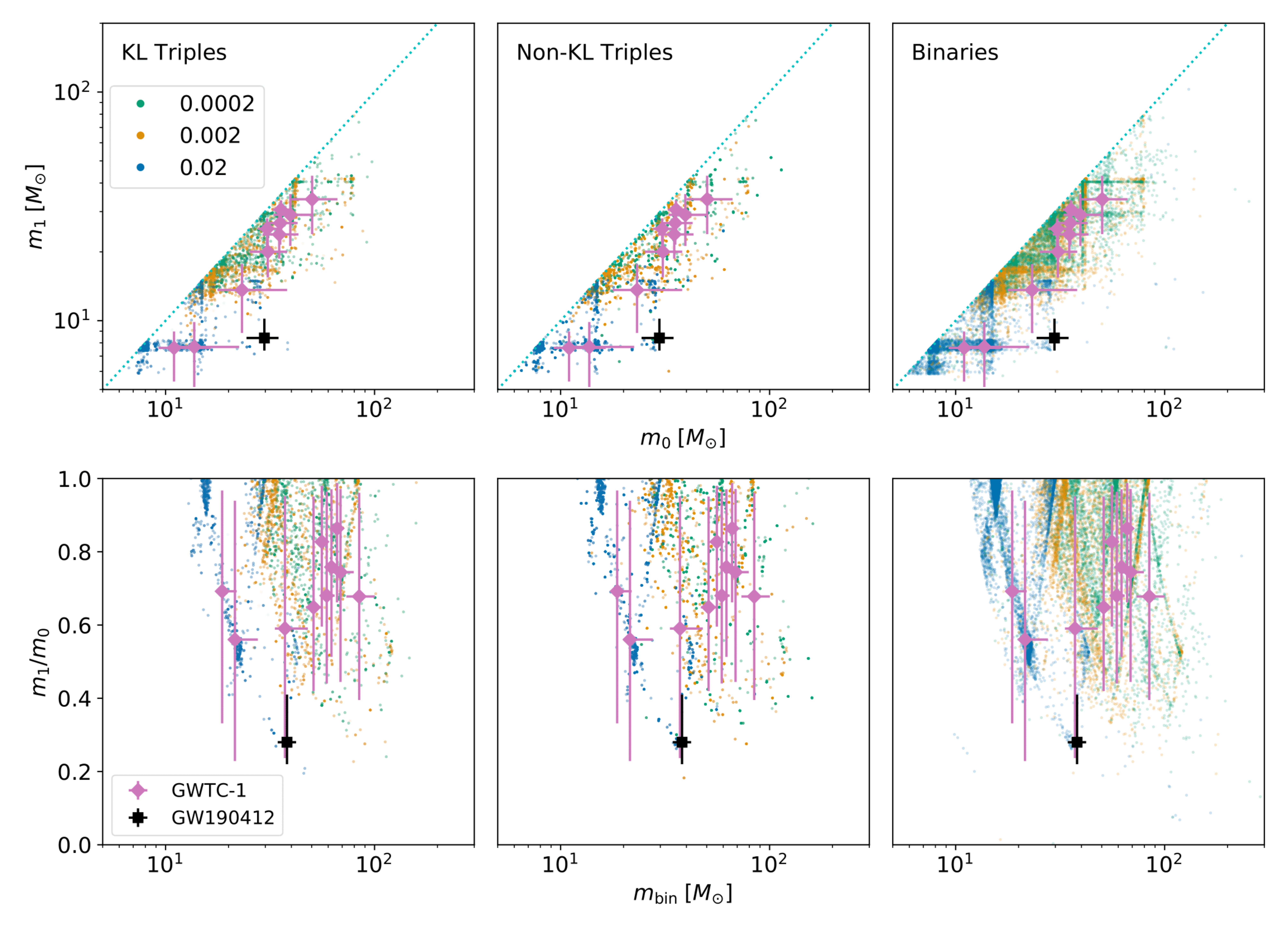}
\caption{Masses and mass ratios of merging binaries colored by metallicity. (Top) Masses of the binary components from this study and \citet{kremer+2019}. The dotted light blue line represents the ratio 1:1. (Bottom) Total binary mass versus the mass ratio. The streaks present in the figures are due to the excess of BHs with mass 40.5 $\msun$, since this is the lower boundary of the (pulsational) pair instability mass gap used in our simulations, and BHs in our simulations typically cannot form at higher masses without direct stellar or BH collisions. Also included are the mergers from the GWTC-1 catalog from O1 and O2 \citep{abbott+2019a,abbott+2019b}, as well as the detection of GW190412 during O3 \citep{LIGO+Virgo2020}. Comparison of the three panels shows minimal differences, again demonstrating that triple evolution has a minimal effect on the mass spectrum.}
\label{fig:masses}
\end{figure*}

In Figure \ref{fig:masseshist}, we compare the total masses, chirp masses, and mass ratios of the merging binaries through both triple and binary channels, where the chirp mass is defined as
\begin{equation}
m_{\rm chirp} \equiv \frac{(m_{0}m_{1})^{3/5}}{(m_0+m_1)^{1/5}}\,.
\end{equation}
In the top panels, we compare the combined (KL and non-KL) triple sample to the binary mergers from \citet{kremer+2019}. We find a peak chirp mass of $\sim 17\msun$ and a peak total mass of $\sim 40\msun$ for the triple mergers. To first order, the mass and mass ratio spectrum between binaries and triples are similar. Only one difference stands out: compared to the binary mergers, mergers in triples show a diminution of low-mass mergers with $\mbin<25\msun$ by approximately $40\%$. This could happen because low-mass binaries are less likely to take part in the binary-binary interactions necessary to produce the triples in our models. In the bottom panels, we compare the KL and non-KL systems to each other and the combined triple sample. Once again, we find minimal differences between these two distributions. This shows the influence of a tertiary companion has a minimal effect on the properties of merging binaries, whereas these properties are determined primarily by the far more frequent binary-mediated dynamical interactions within the cluster core.

We find that our models produce heavier BH mergers when compared to \citet{antonini+2016}, who did not produce any BH mergers with total masses above $\sim 50\msun$, though they find similar peak total and chirp masses. This is most likely due to the wider parameter space for the initial conditions of our GC models. The difference is even more stark when compared to the study of field triples from \citet{antonini+2017}, who were only able to produce binaries in the mass range $13\msun$--$20\msun$, emphasizing the key role of mass segregation in producing low-mass mergers. However, \citet{rodriguez+antonini2018} were able to produce binaries in field triples with a total mass spectrum similar to the one presented here. The difference between this study and earlier studies arises due to a higher limit on the maximum stellar mass in their models. 

In Figure \ref{fig:masses}, we present the component masses and the total mass versus mass ratio for KL mergers, non-KL mergers, and binaries from \citet{kremer+2019} colored by metallicity. For comparison, we also present the properties of the confirmed BBH detections from GWTC-1 and the recent announcement of the detection of GW190412, the first binary BH system with asymmetric masses at high confidence \citep{abbott+2019a,abbott+2019b,LIGO+Virgo2020}\footnote{We do not include GW190814 \citep{LIGO+Virgo2020b} in this comparison because current simulations predict such mergers do not occur at any appreciable rate in GCs \citep{kremer+2019,ye+2020}.}. We find that the three different populations present minimal differences, as expected from the previous discussion. Moreover, we find that the BBH mergers that have been announced so far are consistent with production in a GC, with all the GWTC-1 events that lie within dense regions of both diagrams. On the other hand, GCs are not as efficient at producing events in the mass/mass ratio range as low as that of GW190412 due to the efficiency of mass segregation. More importantly, we find that merger via a hierarchical triple does not increase the percentage of similar events relative to the binary channel.

However, recent works by \citet{rodriguez+2020} and \citet{gerosa+2020} have shown that the relative fraction of low-mass ratio merger events can be enhanced by second- and third-generation mergers in dynamical environments, producing GW190412-like events as a result. Due to the efficiency of mass segregation, the remnant of a previous merger is expected to be twice as massive as the first generation BHs in the cluster \citep[see also][]{samsing+hotokezaka2020}. In \S\ref{subsect:kicks}, we show how triples may contribute to increase the number of second generation mergers.

\subsection{Spins}

\begin{figure}
\includegraphics[width=0.96\textwidth]{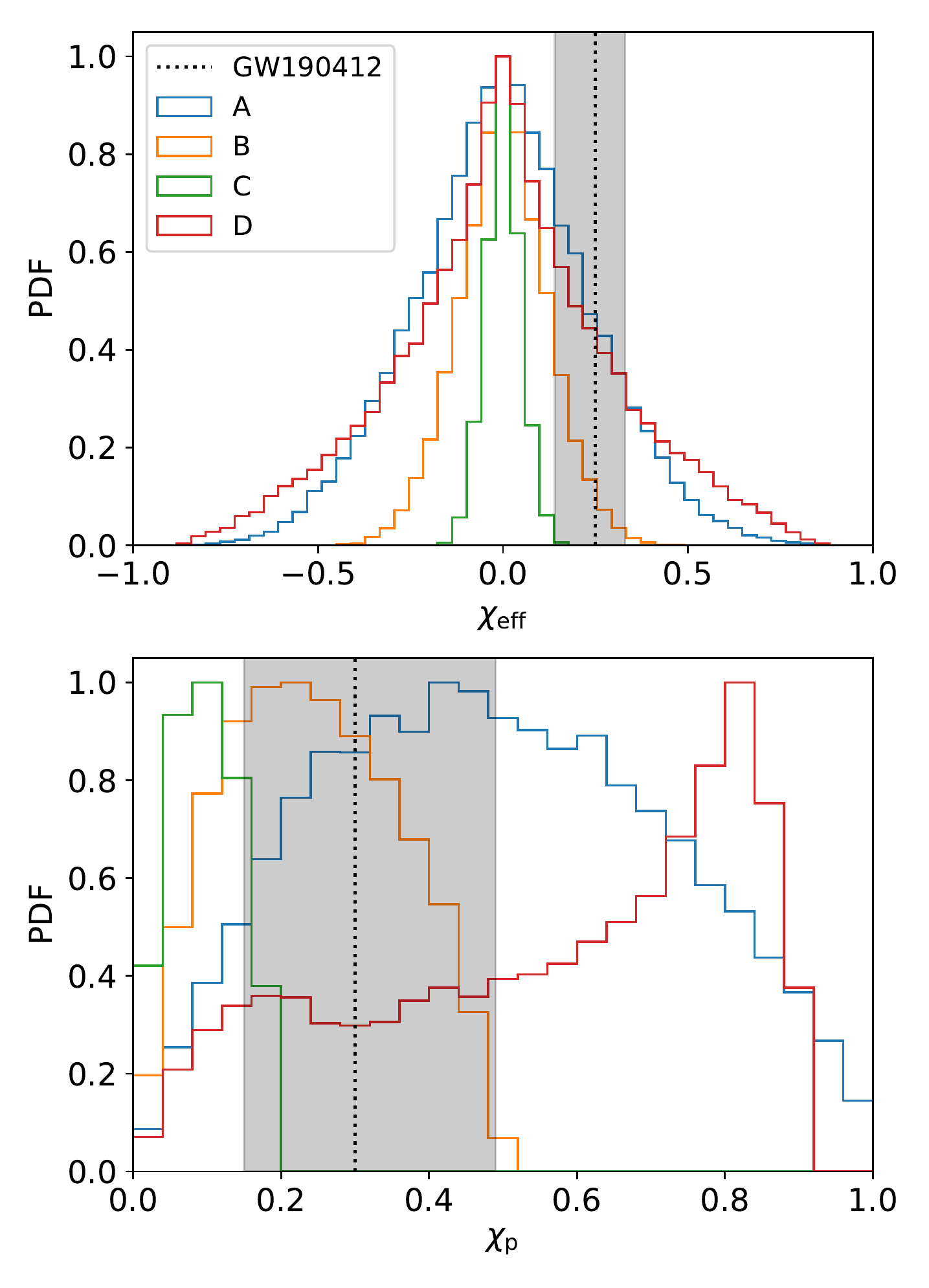}
\caption{Comparison of $\chieff$ (top) and $\chi_{\rm p}$ (bottom) distributions for different spin models from Table \ref{tab:spins}. All distributions are normalized such that their peak has magnitude $1$. Also shown is the $90\%$ probability region for GW190412 \citep{LIGO+Virgo2020} for both quantities.}
\label{fig:chihist}
\end{figure}

\begin{figure*}
\centering
\includegraphics[width=0.65\textwidth]{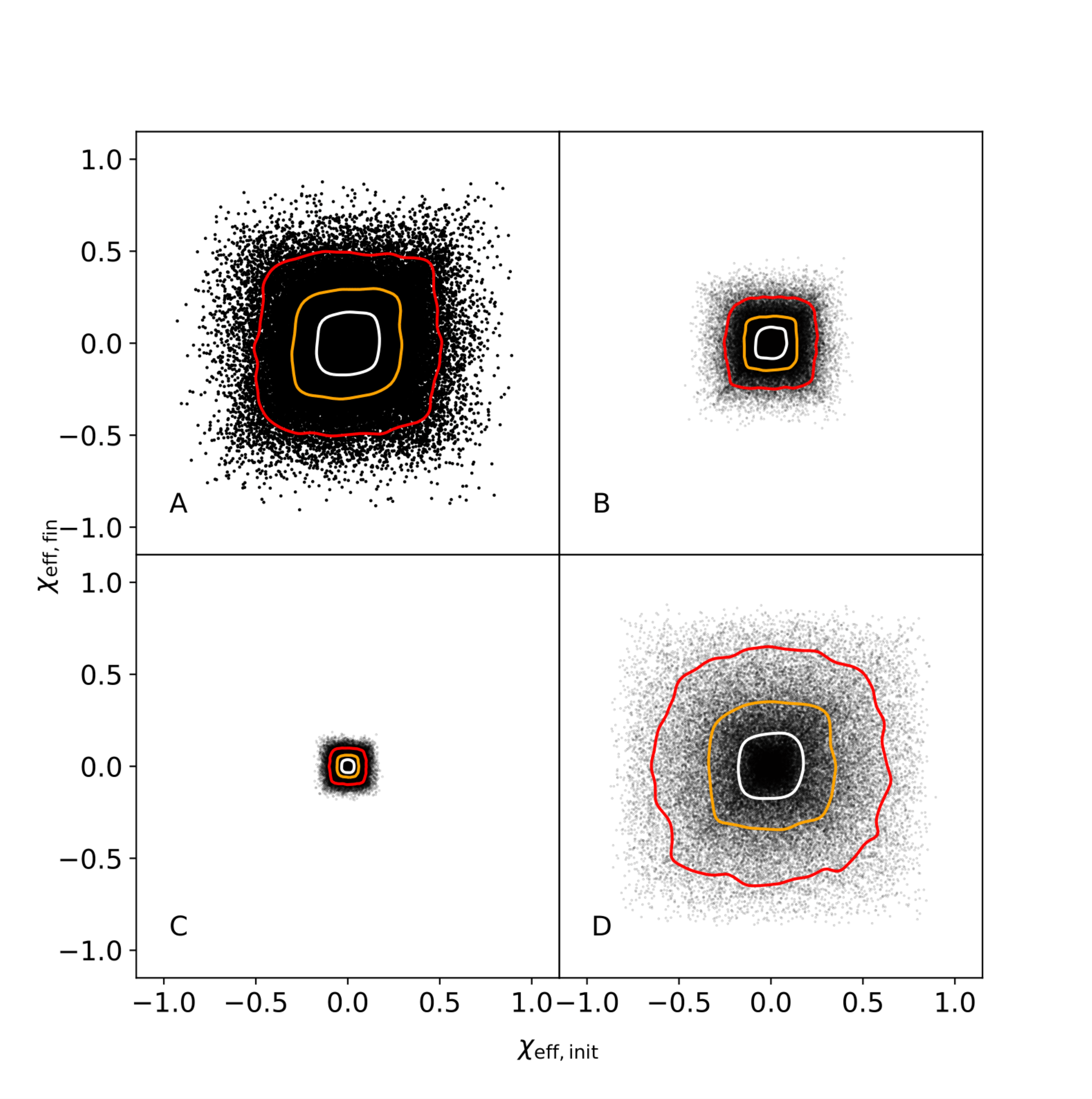}
\caption{Initial versus final $\chieff$ distributions for different spin models. Only KL triples are shown here. Contours show regions containing 30\% (white), 60\% (orange), and 90\% (red) of points. The labels correspond to the model names from Table \ref{tab:spins}. The symmetrical distribution of points in each panel shows that while the overall distribution of misalignment angles remains the same, the individual triples still undergo $\chieff$ evolution.}
\label{fig:spins}
\end{figure*}

\begin{table}
    \centering
    \begin{tabular}{c|c}
        \hline
        Model  &  $\chi$\\
        \hline
        A  &  Uniform in $[0,1)$\\
        B  &  Uniform in $[0,0.5)$\\
        C  &  Uniform in $[0,0.2)$\\
        D  &  Eq. \ref{eqn:belspins}\\
        \hline
    \end{tabular}
    \caption{$\chi$ distributions for different models. All models sample spin vector orientations isotropically.}
    \label{tab:spins}
\end{table}

As the initial distribution of BH spins is unknown, we adopt 4 different models of spin magnitude to bracket uncertainties. The different models are summarized in Table \ref{tab:spins}. For each model, we sample the spin of both inner binary components 10 times each, sampling the orientations isotropically. Models A, B, and C set the natal $\chi$ from a uniform distribution between $0$ and $1$, $0.5$, and $0.2$ respectively. Model D sets the birth spins using the following function which reflects work by \citet{belczynski+2017} set by the BH mass:
\be
\chi = \frac{p_1 - p_2}{2} \tanh\left( p_3 - \frac{m}{\msun} \right) + \frac{p_1 + p_2}{2}
\label{eqn:belspins}
\ee
The quantities in the previous formula have values $p_1 = 0.86 \pm 0.06$, $p_2 = 0.13 \pm 0.13$, and $p_3 = 29.5 \pm 8.5$. Following \citet{gerosa+2018}, we sample values in between the lower and upper limits of the parameters for a BH of a given mass $m$. This last model has the advantage of assigning lower spins to more massive BHs and vice versa. 

We compare the $\chieff$ distributions from different models in the top panel of Figure \ref{fig:chihist}. All models peak around $\chieff\approx 0$, which is to be expected from an isotropic distribution of spin-orbit misalignment angles. This is consistent with the announced detections from O1/O2, which were all consistent with having near zero effective spin. In the absence of GW190412, a model for natal spin such as B or C would be favored if all the events were generated by dynamically-assembled binaries. However, it is extremely unlikely for models B and C to produce a $\chieff$ consistent with that of GW190412. Additionally, as the detection of GW190412 allowed for a well-constrained measurement of $\chi_{\rm p}$, we compare these distributions for different spin models in the bottom panel. Models A and B are strongly favored as they peak within the $90\%$ probability region, while model C is once again strongly disfavored. As such, if we assume that both components of GW190412 are 1G BHs, then a model that allows for higher natal spin is necessary.

We find no significant difference between the initial and final $\chieff$ spin distributions across any of our models, as shown in Figure \ref{fig:spins}. This is in agreement with previous findings from \citet{rodriguez+antonini2018} and \citet{fragione+2019}. The spin vectors in the inner binary evolve in response to two different torques. On the one hand, they are precessing around the inner binary angular momentum according to the relativistic spin-orbit coupling (Eq.~\ref{eqn:spinorbit}). On the other hand, the KL mechanism causes the precession of the inner orbit angular momentum around the outer orbit angular momentum. If the rate of precession from the relativistic spin-orbit coupling is faster than the rate of precession due to KL, then $\chi_{\rm eff}$ evolution is completely suppressed \citep{storch+2014,storch+lai2015,liu+lai2017,liu+lai2018,antonini+2018}. The individual triple systems span the full space of allowed values of initial versus final $\chieff$, even if the initial and final distributions of $\chieff$ do not appear to change (see Fig.~\ref{fig:spins}). As a result, we can safely conclude that spin relativistic precession does not suppress the evolution of $\chieff$ in most cases. In summary, while spin precession does take place in these triples, since the triples are dynamically assembled, the final $\chieff$ distribution will remain isotropic.

Recent work by \citet{belczynski+2020} comparing different stellar evolution models in light of the detections of GWTC-1 has shown that the results from the LIGO--Virgo collaboration strongly favor low natal spins. However, more recent work by \citet{olejak+2020} has shown that GW190412 is consistent with an isolated binary formation scenario, albeit using the older, standard assumptions present in \citet{belczynski+2017}. It is not clear if isolated stellar evolutionary models alone can consistently produce the $\chieff$ distribution in the detected events so far. On the other hand, recent works by \citet{rodriguez+2020} and \citet{gerosa+2020} have shown that it would be possible to reproduce these detections with repeated mergers of previous merger products in extremely large clusters. This would require that BHs are born with low natal spins, so as to retain more of the merger remnants. Further discussion of the importance of spin with respect to retaining merger remnants and producing mergers with second generation black holes will follow in \S\ref{subsect:kicks}. 

Other formation channels lessen the sensitivity to the assumed natal spin model. In particular, previous work on field triples has shown that the spin evolution of these systems can take a wide range of evolutionary paths depending on initial assumptions. While the natal spin assumptions are important, initial spin-orbit misalignment and the properties of the outer orbit play as large a role in determining the final state of the system \citep{liu+lai2019,fragione+2019}. Nevertheless, the field triple channel tends to produce distributions peaked at $\chieff\simeq0$ \citep{antonini+2018,rodriguez+antonini2018}. Another proposed scenario involves a hierarchical $3+1$ quadruple system, wherein the innermost orbit is driven to merger by secular chaotic evolution and then the merger remnant merges with another compact object due to the eKL mechanism \citep{safarzadeh+2020}. While the rates for these events are extremely uncertain, this is another potential formation channel for events like GW190412 \citep{hamers+safarzadeh2020}. Similarly, 2+2 quadruples could also produce GW190412-like events \citep{flr2020}. Another possibility is through the merger of BHs in active galactic nuclei (AGN) disks \citep{tagawa+2018,tagawa+2019,tagawa+2020}. In particular, recent work by \citet{tagawa+2020} has shown that the spin distribution of the O1/O2 events is consistent with mergers in AGN disks, while a hierarchical merger within AGN disks could potentially reproduce the properties of GW190412.

\subsection{Merger Rate}

\begin{figure*}
\includegraphics[width=0.9\textwidth]{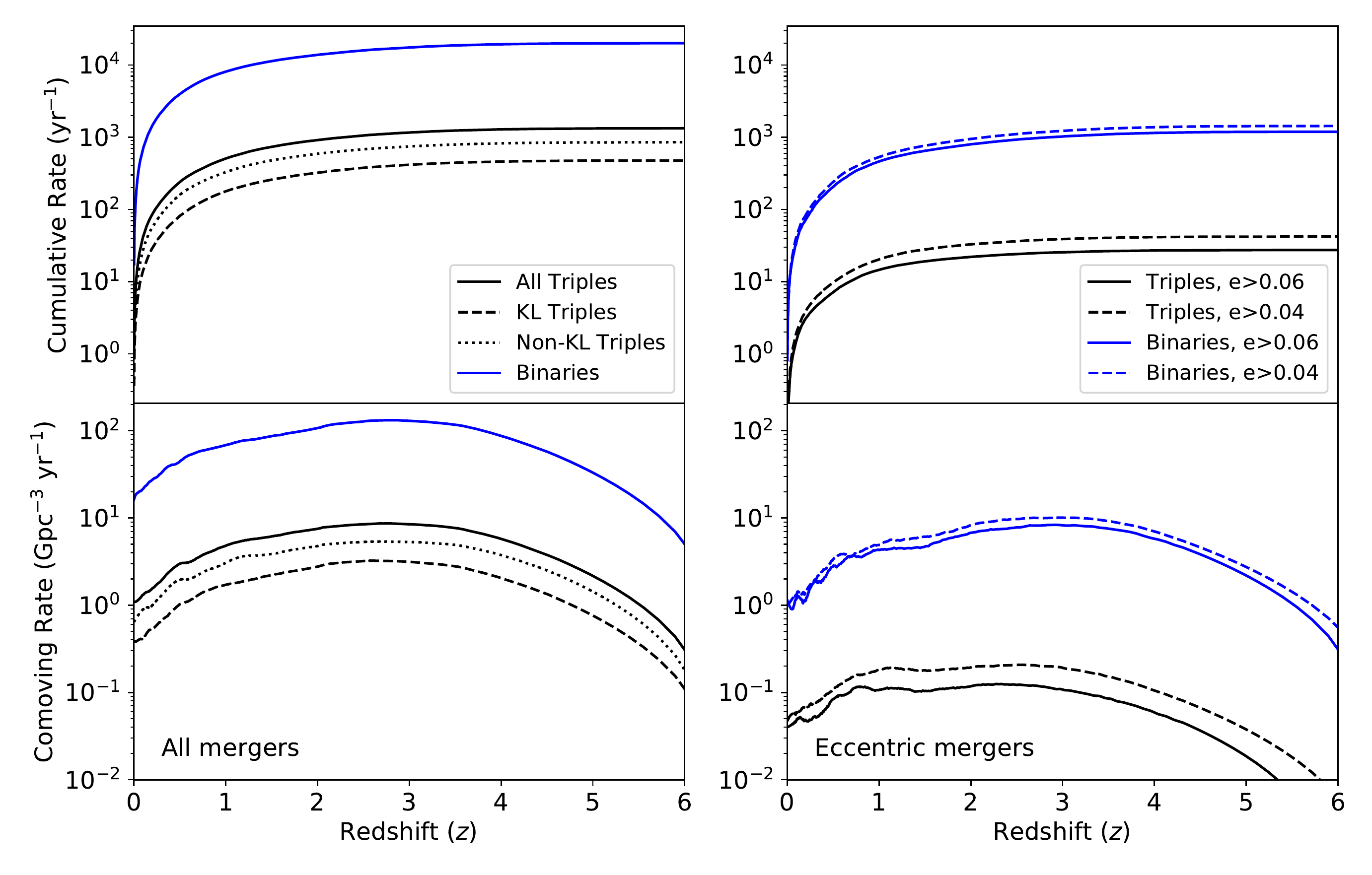}
\caption{Cumulative (top) and comoving (bottom) merger rates for clusters. Blue represents the merger rate of binaries presented in \citet{kremer+2019}. On the left, black lines show the triple mergers of this study. The solid line shows the combined triple rate while the dashed and dotted lines represent KL and non-KL mergers respectively. On the right, black and blue lines represent triple and binary mergers that could have a detectable eccentricity in the LVK band. Since the threshold for detectable eccentricity is mass-dependent, solid and dotted lines are used to bracket the possibilities for different masses.}
\label{fig:times}
\end{figure*}

Following the method of \citet{rodriguez+2015}, the cumulative merger rate is given by
\be
R(z) = \int_0^z \iR(z') \dv{V_c}{z'} \left(1+z'\right)^{-1} \dd{z'}
\ee
Here, $\dv*{V_c}{z'}$ is the comoving volume at redshift $z$ and $\iR(z')$ is the comoving source merger rate, where the comoving rate is given by
\be
\iR=f\times\rho_{GC}\times\dv{N(z)}{t}
\ee
We assume that the volumetric number density of clusters has a constant value of $\rho_{GC}=2.3\,\mathrm{Mpc}^{-3}$ \citep{rodriguez+2015,rodriguez+loeb2018}. $f$ is an $r_v$-dependent scaling factor to incorporate the high-end tail of the cluster mass function not covered in the catalog. Finally, $\dv*{N(z)}{t}$ is the number of mergers per unit time at $z$.

In order to compute $\dv*{N(z)}{t}$, we compile a complete list of $T_{\rm merger}$ for each realization of the merging triples in our sample. For each merger, we resample their merger times by sampling 10 random host cluster ages for each merger and define the effective merger time $t_{\rm effective}=t_{\rm Hubble} - t_{\rm age} + T_{\rm merger}$ using the host cluster ages from the metallicity-dependent age distributions of \citet{el-badry+2019}. We then bin this list of merger times into redshift bins. We must also scale down these rates to correct for oversampling. We divide the rates by a factor of 1000 to account for the resampling with respect to cluster age, triple recoil velocity, and orbital orientation. We must also divide by the total number of cluster models in order to correct for drawing from a large set of cluster models, weighting all models equally for simplicity.

The scaling factor $f$ is included in order to account for the low-mass tail of the cluster mass function not covered by our models (consistent with \citealt{kremer+2019}). In practice, high-mass clusters ($M\gtrsim 5\times10^5\,\msun$) contribute roughly four times the number of mergers compared to low-mass clusters ($M\lesssim 5\times10^5\,\msun$), so $f\approx4$ across different $r_v$ values. Full details can be found in Section 9.2 and Table 5 of \citet{kremer+2019}.

In Figure \ref{fig:times}, we compare the cumulative and comoving merger rates, finding that the triple merger rate remains roughly two orders of magnitude less than the binary merger rate from GCs across all redshifts. Thus, we find a local universe eKL-assisted triple merger rate in GCs $\approx 0.35$ Gpc$^{-3}$yr$^{-1}$. This is consistent with the previous merger rate found by \citet{antonini+2016}. The non-KL merger rate on the other hand is $\approx 0.62$ Gpc$^{-3}$yr$^{-1}$, with a possible range of $\approx 0.2 - 2$ Gpc$^{-3}$yr$^{-1}$. We also show rates for events that may have a detectable eccentricity in the LVK band at design sensitivity \footnote{\citet{gondan+kocsis2019} showed that for the events in the GWTC-1 transient catalog, the minimum detectable eccentricity for low-mass neutron star binaries is $\sim 0.023$ and can reach $\sim 0.081$ for the highest-mass BH binaries. We select the range between $0.04$ and $0.06$ as representative of most of the binaries presented in this study.}. In agreement with the lack of eccentric events thus far \citep{LVCecc2019}, we find low rates $\approx 0.025-0.035$ Gpc$^{-3}$yr$^{-1}$ for the triple channel and $\approx 1$ Gpc$^{-3}$yr$^{-1}$ for the few-body and single-single capture channels. This should be compared to the following merger rate estimates from the LIGO--Virgo collaboration and from studies looking at hierarchical triples in the field: 
\begin{itemize}
    \item LVC rate: $9.7$-$101$ Gpc$^{-3}$yr$^{-1}$ \citep{abbott+2019a,abbott+2019b}
    \item Field Triples: $0.14$-$6$ Gpc$^{-3}$yr$^{-1}$ \citep{silsbee+tremaine2017}; $0.3$-$1.3$ Gpc$^{-3}$yr$^{-1}$ \citep{antonini+2017}; $2$-$25$ Gpc$^{-3}$yr$^{-1}$ \citep{rodriguez+antonini2018}; $0.02$-$24$ Gpc$^{-3}$yr$^{-1}$ \citep{fragione+2019}
    \item Nuclear Star Cluster Triples: $\approx1$-$10^2$ detections yr$^{-1}$\citep{oleary+2009}; $0.6$-$15$ Gpc$^{-3}$yr$^{-1}$ \citep{petrovich+antonini2017}; $1$-$3$ Gpc$^{-3}$yr$^{-1}$ \citep{hoang+2018,fragrish2019,stephan+2019}
\end{itemize}
This rate should be taken as a conservative estimate, as this study only focuses on triples composed of three BHs. We do not analyze the other classes of triples, which may well become BBHs with other companions, as we are unable to self-consistently track the stellar evolution of three separate bodies undergoing the eKL mechanism \citep[see e.g.,][]{dis2020a,dis2020b}.

While it is clear that KL mergers are an addition to the previous merger rate found by \citet{kremer+2019}, it is not clear if all non-KL mergers were included in this previous estimate. In principle, since all of these binaries merged solely due to GW radiation, regardless of whether or not the tertiary companion were present, these events have already been accounted for in the previous binary rate estimate. In \S\ref{subsect:kicks}, we will explain why it is still important to keep track of triple systems whether or not the eKL mechanism dominates the triple evolution.

\subsection{Merger Remnant Retention}
\label{subsect:kicks}

\begin{figure}
\includegraphics[width=0.96\textwidth]{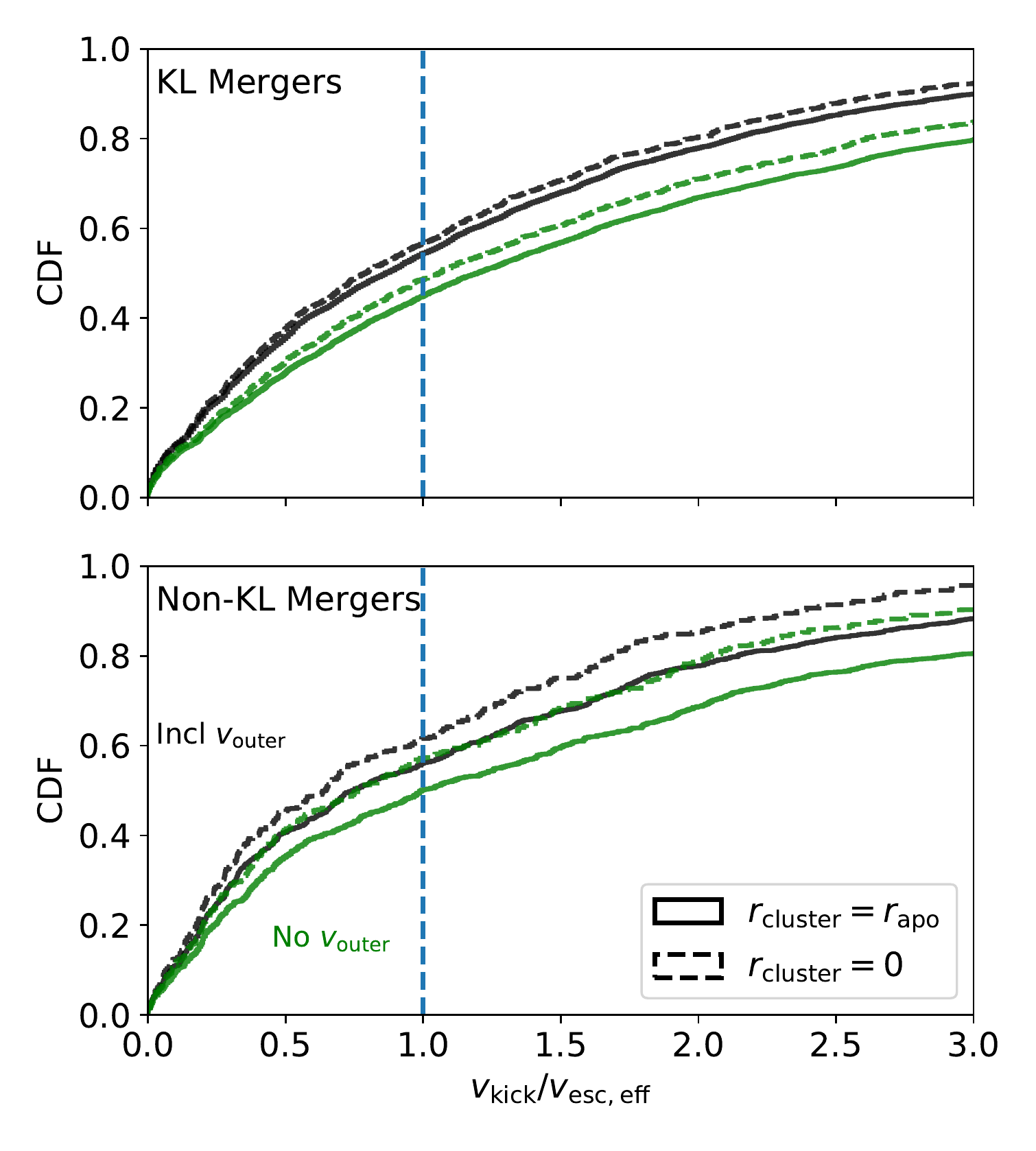}
\caption{Retention fractions of merger remnants for different values of $v_{\rm esc,eff}$ (Eq. \ref{eqn:vesc}) for KL (top) and non-KL (bottom) mergers. Black and green lines correspond to the presence and absence of the tertiary potential. Because the local escape speed depends on the location of the triple in the cluster, we bracket the range of possibilities with the extreme values $r=r_{\rm apo}$ (solid lines) and $r=0$ (dotted lines). While the difference is more apparent for KL mergers due to the compactness of the inner and outer orbits, we find that even for non-KL mergers, the tertiary has an appreciable effect on the retention fraction.}
\label{fig:kicks}
\end{figure}

When two BHs merge, the merger remnant will experience a kick due to the asymmetric emission of GW radiation, depending on the mass asymmetry and the alignment of the spin vectors with the binary orbital plane \citep{campanelli+2007,lousto+zlochower2008,lousto+2010,lousto+2012,lousto+zlochower2013}. The contribution due to the mass asymmetry takes the form
\be
v_{\rm kick} = A \eta^2 \frac{1-q}{1+q} (1+B \eta),
\ee
where $\eta=m_{0}m_{1}/(m_{0}+m{1})^2$, $q$ is the mass ratio, and A and B are numerically-calibrated constants, with $A=1.2\times10^4$ km/s and $B=-0.93$. The numerical relativity simulations from those same studies have shown that merging BHs with even moderate spin greatly increase the magnitude of the kick velocity, depending on their mutual misalignment and the misalignment with the orbital plane, reaching magnitudes of up to thousands of km/s. Because all the models of the \texttt{CMC Catalog} assume that all first generation (1G) BHs are born with no spin \citep{fuller+ma2019}, we elect to only consider the mass asymmetry term in the recoil velocity.

If the kick velocity is greater than the local escape speed, then the remnant will be ejected from the cluster. However, if the merger happens within a triple, then the remnant must also overcome the potential of the tertiary body. As a result, the merger remnant must overcome the local escape speed, which we calculate as follows:
\be
v_{\rm esc,eff} = v_{\rm outer}+v_{\rm esc}-\alpha(r) v_{\rm rec}.
\label{eqn:vesc}
\ee
Here, $v_{\rm esc}$ is the escape speed from the cluster core at the time of merger, $v_{\rm outer}$ is the escape speed from the tertiary companion, and $v_{\rm rec}$ is the recoil kick that the triple experiences at formation. Due to the uncertainty regarding the triple's location within the cluster at the time of merger, we include a factor $\alpha(r) \in [0,1]$ to account for the phase of the orbit and the effect of dynamical friction in reducing the orbit's apocenter \citep{binney+tremaine}. We define this parameter such that $\alpha=1$ if the triple is at the apocenter of its orbit and has experienced no dynamical friction and $\alpha=0$ if the triple is located within the cluster core.

In Figure \ref{fig:kicks}, we compare the retention fraction for both KL and non-KL systems including and not including the potential of the tertiary. Since we do not know the location of the triple within the cluster at the time of the merger, we show the range of possibilities by showing curves where $\alpha=0$ and $\alpha=1$. In reality, the merger retention fraction will be somewhere in between these values. We find that including the tertiary's potential has an appreciable effect on merger remnant retention. We find that for KL mergers, the fraction of retained systems increases by  $\sim 10\%$, whereas the retention fraction increases by  $\sim 6\%$ for non-KL mergers. The effect of the tertiary is much more pronounced for KL-induced mergers since the orbits are more compact. We note that the solid green line in the bottom panel is equivalent to the merger remnant retention fraction for the binary merger channel, as it corresponds to mergers in the cluster core without a tertiary companion. From this, it is clear that triple systems enhance the number of 2G BHs retained in the host cluster.

These results are extremely interesting in light of the recent detection of GW190521 \citep{GW190521a}, a merger event with a total mass of $150 \msun$, whose components could be second-generation BHs. Many studies have discussed the frequency and importance of successive mergers with 2G and higher BHs in massive clusters such as GCs, super star clusters, and nuclear star clusters \citep{rodriguez+2019,antonini+2019,fragsilk2020,gerosa+2020,rodriguez+2020, samsing+hotokezaka2020}. Mergers involving 2G BHs are expected to be more massive, potentially falling within the mass range $\sim 46-133 \, \msun$, the so-called upper mass gap, where no BHs are expected to be formed due to pair instability supernovae \citep{woosley+heger2015,woosley2017,woosley2019}. As a result, if any BH mergers are detected with a component in the upper mass gap, the only way they could have been formed is through prior stellar mergers or successive mergers of BHs \citep{dicarlo+2019b,fragsilk2020,kremer+2020}. This possibility has been recently confirmed with the detection of GW190521, which is consistent with a dynamical origin \citep{GW190521b}. Second, it is expected that a 2G+1G merger will have very asymmetric masses. Mass segregation in globular and higher-mass clusters causes binaries to form primarily between two low-mass members. Since it is unlikely that there will be more than one 2G BH in a cluster at any time, the 2G BH will necessarily form with a 1G BH of much lower mass \citep{rodriguez+2019}. \citet{rodriguez+2019} has shown that, when assuming BHs have no natal spin, 2G+1G mergers typically contribute  $\sim 20\%$ of detectable mergers from GCs while 2G+2G mergers only contribute a few percent. Thus, keeping track of successive mergers is important to understand the true mass and mass ratio spectrum from GCs. It is also worth noting that since the LVK detectors are more sensitive to more massive BHs, mergers involving 2G or higher BHs will make up a larger fraction of the detected BHs.

In addition to the mass gap, \citet{baibhav+2020} has described the existence of the ``spin gap'' in the context of successive mergers. If two BHs are non-spinning at birth, the only way to form a BH where one or both components have high spin is through successive mergers. Even if BHs are born with some natal spin, studies by \citet{berti+volonteri2008,gerosa+berti2017,fishbach+2017} have shown that, regardless of the model of natal spin of 1G BHs, the spin distribution of mergers involving 2G BHs peaks at $\chi\simeq0.7$.  Successive mergers would be the only way to form very high $\chieff$ mergers, requiring the dynamical assembly of binaries. Constraining the rate of higher-generation mergers using observations populating the upper mass gap and the ``spin gap'' will be important to disentangle the isolated field formation channel from the dynamical formation channels. Since 2G+1G and 2G+2G events are not expected to make up a large fraction of detections, it is crucial to accurately constrain the rates of these events.

\subsection{Direct n-body versus Secular Integration}
\label{subsect:n-body}

\begin{figure*}
\includegraphics[width=0.96\textwidth]{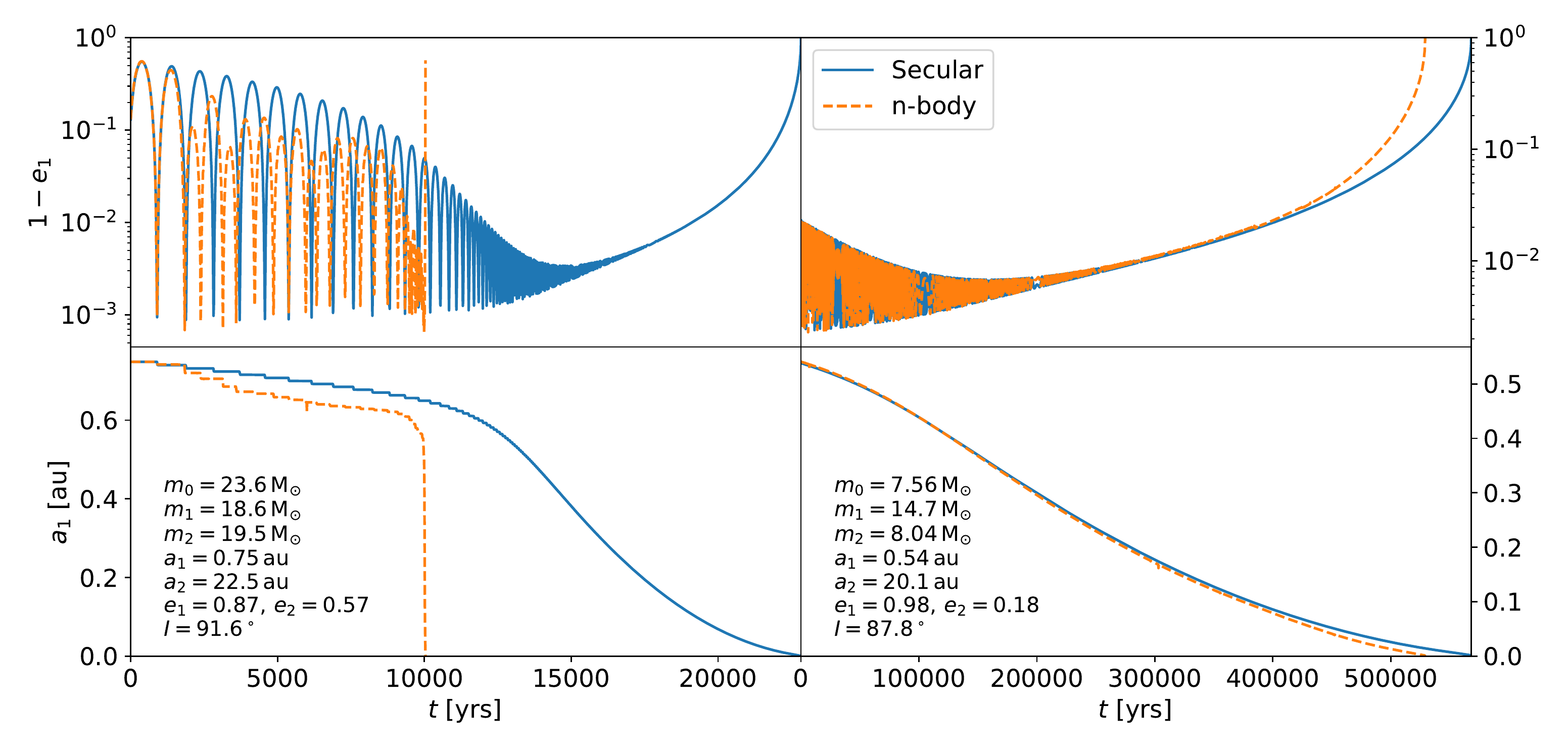}
\caption{Two examples of triple systems from our sample integrated with secular (blue) and direct $n$-body (orange) codes. We show (top) eccentricity and (bottom) semimajor axis evolution of the inner binary. The left example shows a worst case, semi-hierarchical scenario while the right example shows close agreement between the two methods. Less hierarchical systems such as the one shown on the left may merit the use of $n$-body codes to accurately model the system.}
\label{fig:example}
\end{figure*}

At the most extreme eccentricity maxima of the eKL oscillations, the system may enter a dynamical regime wherein the secular approximation breaks down \citep{antonini+perets2012,antonini+2016,fragrish2019}. Furthermore, some systems close to the stability limit may also have angular momentum changes faster than the orbital period of the inner binary \citep{antonini+2014}. In this regime, the secular equations of motion could no longer describe the system correctly, thus necessitating a direct $n$-body integration of the orbits. \citet{antonini+2016} and \citet{fragrish2019} demonstrated that using an $n$-body code instead of a secular code for the same sample increased the total number of mergers and increased the fraction of merging systems entering the LVK frequency band with a finite eccentricity. We confirm these results by simulating a small number ($\sim100$) of the merging systems using a direct integration with \texttt{ARCHAIN}\footnote{We use the version of ARCHAIN available at the following URL: https://arxiv.org/abs/1910.05202} \citep{mikkola+merritt2006,mikkola+merritt2008,archainmanual}. 

In Figure \ref{fig:example}, we show an example of the dynamical evolution of one such system until merger. The system shown on the left is semi-hierarchical with $a_{\rm out}(1-e_{\rm out})/a_{\rm in}\approx12.7$ and the value of the octupole parameter is $\epsilon=0.0034$. Since the system is semi-hierarchical, during certain points in the eKL oscillation, the change in angular momentum of the inner binary happens on a timescale faster than the outer orbital period. This translates to the following numerical condition on the separation between the inner and outer binary \citep[Eq. 9 of][]{antonini+2016}:
\be
\frac{a_{\rm out}(1-e_{\rm out})}{a_{\rm in}} \lesssim \frac{3}{1+e_{\rm out}} \left( \frac{m_2}{\mbin} \right)^{2/3} \left( \frac{\mbin}{\mtrip} \right)^{1/3} \left( \frac{a_{\rm in}}{10^9 \,\mathrm{cm}} \right)^{1/3} .
\label{eqn:secnbod}
\ee
In this case, the condition is satisfied. As a result, the secular integration will not be able to reproduce the most extreme peaks of the eccentricity oscillations. In both the secular and the $n$-body integration, with each peak in the eccentricity oscillation, the semimajor axis decreases by a small but appreciable amount. However, while the two different numerical methods show agreement during the first two eKL oscillations, the center panel shows that the semimajor axis begins decreasing more rapidly in the $n$-body integration, implying a larger peak in eccentricity compared to the secular integration. Each successive eccentricity oscillation accelerates the divergence of the results between these two methods. While the secular integration shows a gradual decline in $a_{\rm in}$ as GW radiation begins to dominate, the direct $n$-body integration shows the inner binary merges in about half the time compared to the secular integration. Furthermore, we find direct integration indicates an eccentricity of $\sim 2\times 10^{-3}$ at $10 \,\mathrm{Hz}$ versus $\sim 7\times 10^{-4}$ from the secular integration.

On the other hand, the system shown on the right is more hierarchical, with $a_{\rm out}(1-e_{\rm out})/a_{\rm in}\approx30.5$ and Eq.~\ref{eqn:secnbod} is not satisfied. As such, there is much better agreement between the two numerical schemes.

To summarize, using an $n$-body integration method instead of a secular integration method increases the total number of mergers and the fraction of merging systems entering the LVK frequency band with a high eccentricity \citep[e.g.,][]{antonini+perets2012,antonini+2016,fragrish2019}, which we have confirmed by simulating a small sample of our triple systems. In light of this and also due to the large computational cost from the number of systems requiring integration, we choose to primarily restrict ourselves to using a secular code. Finally, we note that there some issues may arise with incorporating the conservative post-Newtonian effects into $n$-body simulations, since the Newtonian energy is no longer explicitly conserved in this approximation \citep[see e.g. Fig.~2 of ][]{rodriguez+2018}. As a result, the incorporation of the post-Newtonian effects could introduce an increase in energy during the pericenter passage. In hierarchical triples, this could cause the triple to move further into the chaotic regime and increase the amount of GW emission, possibly changing the inspiral waveform.

\section{Conclusions}
\label{sect:conc}

As hundreds of GW detections are expected in the coming years, it is important to understand the unique properties each merger channel imprints on the merging binary population. In this study, we examined the role of hierarchical triples in GCs and their contribution to the population of merging BBHs using the \texttt{CMC Cluster Catalog} from \citet{kremer+2019}. We studied the properties of the merging sample that could be used to discriminate between this and other merger channels, in particular focusing on the eccentricity, masses, and spins compared to binary mergers from the same cluster models by numerically integrating the secular equations of motion of the triples.

We find a conservative local universe merger rate of $\sim 0.35$ Gpc$^{-3}$yr$^{-1}$ for KL mergers, compared to a binary merger rate of $\sim 20$ Gpc$^{-3}$yr$^{-1}$ from the same cluster models. We also find that $\sim 3\%$ of the previously reported binary merger rate from GCs may take place in non-KL triple systems.

In agreement with previous studies, we find that eKL-induced mergers exhibit high eccentricities, which may be detectable by LVK at design sensitivity. We once again emphasize that much previous work has been done to show that numerical methods relying on the secular approximation underpredict the eccentricity of merging systems \citep{antonini+2016,grishin+2018,fragione+2019trip}. Thus, while our sample shows that only a small number of systems would be eccentric within the LVK band, we expect that many more systems would merge with detectable eccentricity than our results suggest. In addition, we show that a significant fraction of the triples produced by GCs will have high eccentricity in the LISA frequency range. Works by \citet{randall+xianyu2019,hoang+2019,deme+2020,emami+loeb2020} have shown the potential for LISA to be able to detect the peaks of the eKL oscillation. We also find that, since GW radiation is efficient in reducing the eccentricity of merging binaries, sub-Hz detectors such as the proposed DECIGO detector \citep{kawamura+2011} will be crucial to understand the contribution of GCs to the merger rate \citep[see also][]{samsing+2019}. Note, that due to limitations of the \texttt{CMC} code, we are not including effects from weak secular interactions that otherwise have been shown to both increase the BBH merger rate and give rise to dynamics similar to those of eKL interactions, including orbital spin-flips, and eccentricity excitations \citep{samsing+2019b,hamers+samsing2019,hamers+samsing2020}.

We find very close agreement between the mass and mass-ratio distributions of binaries merging with and without triple companions, demonstrating that these distributions are primarily shaped by the binary-mediated dynamical interactions within the cluster. Similarly, we find that, since the binaries and triples are dynamically assembled, the effective spin distribution from this channel is almost entirely determined by the natal spin function. As a result, mass and spin are not smoking guns of a triple-induced merger in a dynamical environment.

We find that mergers within triple systems will enhance the number of retained second generation BHs in GCs as long as first generation BHs are born with negligible spin \citep{fuller+ma2019}. More importantly, this result holds for all mergers in triples, whether or not the binary was driven to merge by the eKL mechanism. \citet{baibhav+2020} has shown that successive mergers are a key distinction of dynamically-assembled merging binaries. 

Hierarchical triples are expected to be an important channel for producing LVK sources, as well as sources for future detectors. Our results show that while triples contribute a small fraction of mergers from GCs, the detection of even a single triple merger will be very significant due to its probability of having a detectable eccentricity in the LVK frequency band. We leave it to a future study to see how sensitive these results are to different assumptions such as the primordial binary fraction or the value of the hard-soft boundary. Furthermore, as we have shown that triple mergers can potentially increase the number of retained second-generation BHs, we hope to successfully implement a self-consistent treatment of triples in \texttt{CMC}.

\section*{Acknowledgements}

We thank Fabio Antonini and Nathan Leigh for useful discussions. Our work was supported by NSF Grant AST-1716762. Computations were supported in part through the resources and staff contributions provided for the Quest high performance computing facility at Northwestern University, which is jointly supported by the Office of the Provost, the Office for Research, and Northwestern University Information Technology. This work also used computing resources at CIERA funded by NSF Grant PHY-1726951 and computing resources provided by Northwestern University and the Center for Interdisciplinary Exploration and Research in Astrophysics (CIERA). GF acknowledges support from a CIERA Fellowship at Northwestern University. KK acknowledges support from an NSF Astronomy and Astrophysics Postdoctoral Fellowship under award AST-2001751. SC acknowledges support from the Department of Atomic Energy, Government of India, under project no. 12-R\&D-TFR-5.02-0200. JS acknowledges support from the European Unions Horizon 2020 research and innovation programme under the Marie Sklodowska-Curie grant agreement No. 844629. S.N. acknowledge the partial support of NASA grant No.~80NSSC19K0321 and No.~80NSSC20K0505. S.N. also thanks Howard and Astrid Preston for their generous support.

\software{\texttt{Kozai} \citep{rodriguez+antonini2018,rodriguez+2018},\texttt{OSPE} \citep{naoz+2013a}, ARWV 1.7 \citep{mikkola+merritt2006,mikkola+merritt2008,archainmanual}}

\bibliographystyle{yahapj}
\bibliography{main}

\end{document}